\documentclass[article]{revtex4}

\usepackage{graphics}
\usepackage{bm}

\newcommand{\ket}[1]{| #1 \rangle}
\newcommand{\parenref}[1]{(\ref{#1})}
\newcommand{\msolar}{\ensuremath{M_{\odot}}}
\newcommand{\lps}{\ensuremath{\lambda'}'s\ }
\def\vev#1{\left\langle #1 \right\rangle}
\newcommand{\me}[3]{\ensuremath{\left\langle{#1}\vphantom{#2 #3}
\right|{#2}\left|\vphantom{#1 #2}{#3}\right\rangle}}
\newcommand{\bea}{\begin{eqnarray}}
\newcommand{\eea}{\end{eqnarray}}
\newcommand{\beq}{\begin{equation}}
\newcommand{\eeq}{\end{equation}}
\newcommand{\bay}{\begin{array}}
\newcommand{\eay}{\end{array}} 
\newcommand{\Pm}{(1-\g_5)}
\newcommand{\g}{\gamma}
\newcommand{\SW}{\sin^2\theta_w}

\begin{document}

%\prepint{UCSD/04/AAA/000}

%TITLE
\title{Flavor Changing Supersymmetry Interactions in a Supernova}

\author{Philip S. Amanik, George M. Fuller, Benjamin Grinstein}
\affiliation{Department of Physics, University of California, San Diego, La Jolla, CA 92093-0319}

\date{\today}

\begin{abstract}
We consider for the first time R-parity violating interactions of the Minimal Standard Supersymmetric Model  involving neutrinos and quarks (``flavor changing neutral currents'', FCNC's) in the infall stage of stellar collapse. Our considerations extend to other kinds of flavor changing neutrino reactions as well.   We examine non-forward neutrino scattering processes on heavy nuclei and free nucleons in the supernova core.  This investigation  has led to four principal original discoveries/products: (1) first calculation of neutrino flavor changing  cross sections for spin $1/2$ (e.g. free nucleon) and spin $0$ nuclear targets; (2) discovery of nuclear mass number squared $(A^2)$ coherent amplification of neutrino-quark FCNC's; (3) analysis of FCNC-induced alteration of electron capture and weak/nuclear equilibrium in the collapsing core; and (4) generalization of the calculated cross sections (mentioned in 1) for the case of hot heavy nuclei to be used in collapse/supernova and neutrino transport simulations.  The scattering processes that we consider allow electron neutrinos to change flavor during core collapse, thereby opening holes in the $\nu_e$ sea, which allows electron capture to proceed and results in a lower core electron fraction $Y_e$.  A lower $Y_e$ implies a lower homologous core mass, a lower shock energy, and a greater nuclear photo-disintegration burden for the shock.  In addition, unlike the standard supernova model, the core now could have net muon and/or tau lepton numbers.  These effects could be significant even for supersymmetric couplings below current experimental bounds.  
\end{abstract}

\maketitle

%ENDTITLE

\section{Introduction}
In this paper we consider the effects of certain Supersymmetry (SUSY) interactions during the core collapse and explosion stages of a supernova. These interactions violate lepton number or baryon number and the particular ones we are concerned with allow neutrinos and/or antineutrinos  to change flavor by scattering off $d$ quarks.  We will discuss the consequences that such flavor changing interactions might have for current core collapse supernova models.  
We note that our considerations apply to neutrino flavor changing interactions in general, not just those arising from SUSY  models.  

As the understanding of core collapse supernovae is far from complete, constraints on these SUSY couplings, or couplings for flavor changing neutrino interactions in general, do not follow from our considerations.  However, we do show how such interactions can result in significant alterations in the current model for supernova core collapse and explosion.  It is conceivable that a better understanding of the supernova phenomena, or the detection of a galactic supernova neutrino signal, could allow the effects discussed in this paper to be turned into hard constraints, and/or to be used as signatures for R-parity violating SUSY.  Likewise the same conclusions would apply to other nonstandard flavor changing neutrino interactions. Here, by hard constraints we mean those that can be taken with the same reliability and confidence level as constraints derived from terrestrial and accelerator-based experiments.

R-parity violating interactions have previously been investigated in the context of coherent forward scattering in the case of the  sun\cite{GMPsol,Rsol,BPWsol,GPsol,Bsol,KBsol} and for the late stages of supernovae, r-process nucleosynthesis, and the associated neutrino signal\cite{NRVfcncsn,MKfcncsn,FLMMfcncsn}.  
These papers considered quantum mechanical oscillations between neutrino flavor eigenstates.  
The approach in these papers was similar to the Mikheyev-Smirnov-Wolfenstein mechanism\cite{MSW} for matter enhanced neutrino mixing due to Standard Model (SM) interactions.  

By contrast, in this paper we look at the interactions in an entirely different context. In particular, we examine elastic and inelastic non-forward neutrino flavor changing scattering during the core infall epoch and immediately after core bounce.   Neutrinos become 
trapped in the core during infall because of the large number of scatterings they have on heavy nuclei.  The cross section
for elastic coherent scattering of neutrinos on heavy nuclei via the SM weak neutral current (NC) is large compared to the cross sections for other types of scattering.  Hence, the number of NC nuclear scatterings is larger than the number of other types of scatterings. It is this type of scattering which is principally responsible for the neutrinos having transport mean free paths which are small compared to the size of the core. This regime corresponds to \lq\lq neutrino trapping.\rq\rq\ 

The SUSY interactions we consider behave in an analogous fashion to the SM NC neutrino scattering case.  In particular, the flavor changing SUSY interactions allow for elastic coherent scattering of neutrinos on heavy nuclei.  We will show that the cross section for this process has much the same form as the cross section for the SM neutral current process.  However, there are two important differences between these cross sections.  The first is that the SUSY cross sections are smaller than the SM cross sections because the SUSY interactions are weaker\cite{Chemtob, Barbier,BPWsol} than the SM interactions.  The second difference is that the SUSY interactions allow neutrinos to change flavor.   For example, an electron neutrino can change into a muon or tau neutrino by scattering off a nucleus.  

We point out in this paper that even though the interactions are weaker, non forward SUSY scattering events are still important because of the large number of scatterings neutrinos undergo in the stellar collapse environment 
and because these interactions allow for flavor changing.  This flavor \emph{changing} is different than the flavor \emph{transformation} (that arises from quantum mechanical neutrino oscillations) which has been studied in the works mentioned above.  The number of scatterings is large in the post trapping collapse stage because of  coherent scattering on heavy nuclei.  Post bounce, the number of scatterings is large in the neutron star because the density is so high.   

The effects of changing neutrino flavor in the infall epoch of supernova core collapse has been treated in 
Ref. \cite{fmws}. 
In that work, coherent matter- and neutrino background-enhanced active-active neutrino flavor mixing/transformation during the infall epoch was examined.  General nonstandard neutrino effects and nonstandard neutrino interactions in supernova core collapse are discussed in Ref. \cite{fmw} 

Allowing $\nu_e$'s to change flavor can alter the electron capture and thermal physics in the core.
In fact, the dynamics of core collapse is sensitive to the electron fraction because relativistically degenerate electrons provide most ($> 90\%$) of the pressure in the core.  
By allowing electron neutrinos to change flavor, holes are opened in the Fermi-Dirac sea of $\nu_e$'s.  These holes allow the electron capture reaction \mbox{$e^- + p \leftrightarrow n + \nu_e$} to proceed to the right.  The entropy of the core, homologous core mass, and the initial shock energy at core bounce all depend on the number of electrons in the core during collapse.  We will review how decreasing $Y_e$  can change these quantities, and what implications such changes might have for the supernova model as a whole.  We  perform a simple analysis to estimate the decrease in $Y_e$ during infall due to the SUSY flavor changing interactions. 

Another obvious consequence of allowing flavor changing neutrino interactions to proceed during infall is that seas of $\nu_\mu$'s and/or $\nu_\tau$'s with net lepton numbers can be built up. This is in stark contrast to the standard supernova model where at core bounce and subsequent to it, there are zero net mu and tau lepton numbers, but a significant electron lepton fraction of about $Y_L =Y_e+Y_{\nu_e} \approx 0.35$. An initial lepton number in mu and tau neutrinos could alter all of the subsequent neutrino evolution history in the core and could alter the neutrino signal as well.  Lowering the trapped electron lepton number fraction post bounce also results in an altered equation of state and changes in neutrino transport.   We will discuss incorporating these interactions in numerical simulations in order to account for feedback on the system and get a more detailed picture of the effects throughout all these regimes.  

The paper is organized as follows.  Sec. II will be devoted to SUSY and a derivation of the interactions we are considering.  In Sec. III we discuss the nuclear physics of these interactions and derive relevant cross sections.  In Sec. IV we discuss the supernova model and give our analysis.  In Sec. V we give conclusions.

\section{Supersymmetry}
Supersymmetry\cite{susy,WBsusy,BLsusy,Msusy} is a proposed symmetry of particle physics.  Stated simply, its main feature is that each particle has a partner with opposite spin statistics. Stated another way, each fermion particle has a boson partner, and each boson has a fermion partner.  There are many different models involving SUSY, each containing different numbers of unknown parameters.  If SUSY is indeed a symmetry of nature, it has not been detected yet.  

Supersymmetry is attractive for theoretical and computational reasons.  For example, the underlying symmetry group of the SM is the Lorentz group.  The underlying symmetry group of SUSY is obtained by adding a generator (which represents the super-partner particles) to the generators of the Lorentz group. Thus the underlying  symmetry group of SUSY is a natural and simple extension of the Lorentz  group making SUSY a natural and simple extension to the SM.  
As another example, the three coupling constants of the SM are not equal at low energies, and remain distinct when computed at higher energies.   In SUSY models, these couplings are still distinct at low energies but become equal when calculated at higher energies.  In other words, SUSY is a candidate for a Grand Unified Theory.  Finally, some SM calculations lead to infinities and additional techniques are required to deal with them.  However, the same calculations in SUSY do not lead to infinities because contributions from particles and their partners cancel each other (due to minus signs arising from opposite spin statistics).  It is a feature of SUSY that these infinities are not present.  For this and other reasons, SUSY is taken quite seriously as an extension to the SM.  

The SUSY model we are interested in is the Minimal Standard Supersymmetric Model (MSSM) \cite{Msusy,JGmssm} with additional interactions which break R-parity\cite{rparity,BGHsusy}. 
The MSSM contains the minimum number of fields to describe the known \footnote{Right handed neutrinos (and their superpartners) are excluded from the MSSM.} SM particles and their superpartners.  R-parity is an additional discrete symmetry which requires that an interaction must have an even number of SUSY particles.  If, for example, R-parity is conserved, then the lightest supersymmetric particle (LSP) would not be allowed to decay.   This is prevented because the LSP decay would involve a vertex with two SM model particles but only one SUSY particle.  R-parity violating interactions also violate lepton number (L) or baryon number (B).  

The R-parity violating interactions we consider come from adding the following L violating terms to the superpotential\footnote{A concise discussion of these and other terms which can be added to the superpotential, and implications of adding all these terms, is given in Ref. \cite{Herczeg}.}:
\begin{eqnarray}
&\lambda_{ijk}L_iL_jE^c_k& \label{lepsf} \\
&\lambda'_{ijk}L_iQ_jD_k^c&.\label{qrksf}
\end{eqnarray}
Here $L_i$ and $Q_i$ are lepton and quark $SU(2)$ doublet  superfields, $E^c_i$ and $D^c_i$ are lepton and quark singlet superfields, and the $\lambda$'s are coupling constants.  (We will later specialize our discussion by considering these couplings to be real.)  The Roman subscripts are family indices and there is an implicit contraction of $SU(2)$ indices of the doublet superfields.  Note that in Eq. (\ref{lepsf}) $\lambda_{ijk}$ is antisymmetric in $i$ and $j$.  

In general, a superfield consists of scalar, fermion or vector fields and is a function of spacetime coordinates and so-called superspace coordinates.  The scalar and fermion component fields of a particular superfield represent a particle and its super-partner.  The superspace coordinates  appear as anticommuting Grassman variables.  A superfield can be expanded in terms of its component fields and these Grassman variables.  Likewise, products of superfields such as those that appear in Eq.s (\ref{lepsf}) and (\ref{qrksf}) can be expanded resulting in products of individual particle and sparticle fields.   Kinetic and potential terms can be formed from products of superfields and these can also be expanded in terms of component fields.  A supersymmetric Lagrangian is obtained through a procedure where kinetic, potential and any additional interaction terms are expanded and then the superspace coordinates of the result are integrated over\footnote{For more details on superfields and constructing a Lagrangian from a superfield action we refer the reader to Ref.s \cite{WBsusy,BLsusy,Msusy}.  For a quicker review we recommend Ref. \cite{JGmssm}.}.   

Applying this procedure to Eq.s (\ref{lepsf}) and (\ref{qrksf}) gives the following interactions in the Lagrangian\cite{BGHsusy}:
\begin{eqnarray}
\mathcal{L}=\lambda_{ijk}[\tilde{\nu}^i_L\bar{e}^k_Re^j_L + \tilde{e}^j_L\bar{e}^k_R\nu^i_L + 
(\tilde{e}^k_R)^*\overline{\nu^c}^i_Le^j_L  \nonumber \\
- (i \leftrightarrow j)] + H.c.  \label{leplag} \\
 \mathcal{L}=\lambda'_{ijk}[\tilde{\nu}^i_L\bar{d}^k_Rd^j_L + \tilde{d}^j_L\bar{d}^k_R\nu^i_L + 
 (\tilde{d}^k_R)^*\overline{\nu^c}^i_Ld^j_L \nonumber \\
 -\tilde{e}^i_L\bar{d}^k_Ru^j_L - \tilde{u}^j_L\bar{d}^k_Re^i_L + 
 (\tilde{d}^k_R)^*\overline{e^c}^i_Lu^j_L] \nonumber \\
 + H.c. \label{qrklag}
\end{eqnarray}
Here, sparticle scalar fields are denoted by a tilde.  For example, $\tilde{d}^j_L,\nu^i_L,\bar{e}^k_R$ are left handed down-type squark, left handed neutrino type fermion, and right handed electron type fermion fields, repsectively.  Note that where two left handed fields are contracted, the first field is charge conjugated.
The interactions we are considering in this paper come from Eq. (\ref{qrklag}).  In particular, the second and third terms in the first line of Eq. (\ref{qrklag}) each involve a neutrino, down type quark and down type squark.  These terms and their Hermitian conjugates give flavor changing neutrino scattering with $d$ quarks at the tree level. The vertices of these four terms are given in Table \ref{table 1}, where for illustrative purposes, we have taken the couplings to be real.

\begin{figure}
\resizebox{8cm}{!}{\includegraphics{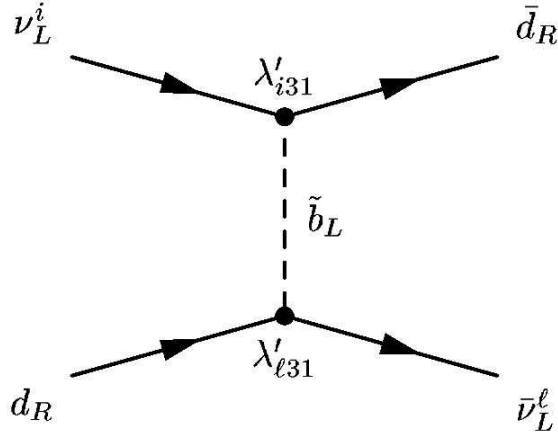}}
\caption{Flavor changing neutrino scattering with a $d$ quark. This tree level diagram is obtained by joining the first two vertices in Table \ref{table 1}.  Time advances from left to right and the labels are for fermion fields.}
\label{figscatt}
\end{figure}

\begin{figure}
\resizebox{8cm}{!}{\includegraphics{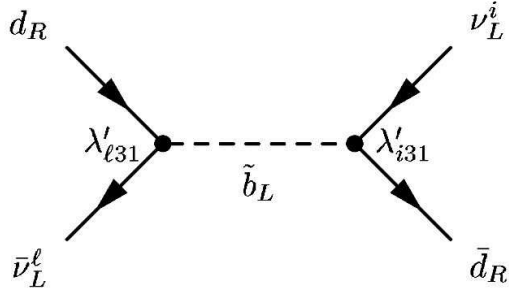}}
\caption{Flavor changing anti-neutrino scattering with a $d$ quark. This tree level diagram is obtained by joining the first two vertices in Table \ref{table 1}.  Time advances from left to right and the labels are for fermion fields.}
\label{figantiscatt}
\end{figure}

We can form tree level diagrams out of the first two vertices in Table \ref{table 1}.  Such diagrams are given in Fig.s (\ref{figscatt}) and (\ref{figantiscatt}), where we have chosen exchange of a $b$ squark.  Since the indices are unspecified we can change neutrino flavor by choosing different indices for the initial and final neutrinos.  In the limit of low energy scattering, we can neglect the squark momentum in its propagator and then the amplitude for the process in Fig.\ (\ref{figscatt}) is
\begin{eqnarray}
\mbox{\Large \textit a} &=& \frac{\lambda'_{i31}\lambda'_{\ell31}}{m^2_{\tilde{b_L}}}\bar{d}_R\nu^i_L\bar{\nu}^\ell_Ld_R   
\nonumber \\
&=&-\frac{\lambda'_{i31}\lambda'_{\ell31}}{2m^2_{\tilde{b_L}}}\bar{\nu}^\ell_L\gamma^\mu\nu^i_L\bar{d}_R\gamma_\mu d_R ,
\end{eqnarray}
where  the second line has been Fierz transformed.  The general low energy effective Lagrangian for our interactions can likewise be formed and is given by\cite{Rsol}
\begin{eqnarray}
\mathcal{L}_{\textrm{\scriptsize eff}}=\frac{\lambda'_{ijk}\lambda'_{\ell mk}}{2m^2_{\tilde{d}^k_R}}
\bar{\nu}^\ell_L\gamma^\mu\nu^i_L\bar{d}^m_L\gamma_\mu d^j_L 
\nonumber \\
-\frac{\lambda'_{ijk}\lambda'_{\ell jn}}{2m^2_{\tilde{d}^j_L}}
\bar{\nu}^\ell_L\gamma^\mu\nu^i_L\bar{d}^k_R\gamma_\mu d^n_R .  \label{fclag1}
\end{eqnarray} 
The $\lambda'$ coupling constants appearing in these interactions have been constrained with the assumption of 100 GeV as a lower bound for all squark masses\footnote{This has been done in Ref. \cite{BGHsusy} and for a more recent review see Ref. \cite{Chemtob}.}.  

We are really interested in products of $\lambda'$'s such as those that appear in Eq. (\ref{fclag1}).  In some cases, limits on individual $\lambda'$'s are used to constrain products of $\lambda'$'s while in other cases particular products of $\lambda'$'s are constrained.  The products of $\lambda'$'s for the flavor changing neutrino scatterings we are interested in are constrained to be less than the range
$10^{-2} -10^{-5}$\cite{BPWsol,Chemtob,Barbier}. 

The Lagrangian in Eq. (\ref{fclag1}) gives neutrino interactions with quarks.  In the next section we derive the cross sections for neutrino scattering with nuclei and free nucleons.
 
\section{Nuclear Physics}
In this section we fold the SUSY leptonic and hadronic currents of the Lagrangian in Eq. \parenref{fclag1} into the physics of nuclei in a hot medium. In particular, we discuss the  coherent elastic scattering cross sections for nuclear target states with angular momentum $0$ and $1/2$ and we present detailed forms for their low momentum transfer limits.  We also preface this discussion with a brief exposition of the physics of the inelastic SUSY flavor changing (FC) interaction channels involving nuclei. 

As pointed out by Bethe, Brown, and co-workers \cite{BBAL} over twenty years ago, the salient feature of the gravitational collapse infall epoch of Type II supernova progenitors is that the entropy is low, and \emph{remains} low throughout collapse.   In units of Boltzmann's constant $k$, the entropy per baryon is $s\sim 1$, some ten times lower than it is in the center of the sun.  An immediate consequence of this is that nucleons will tend to reside inside large nuclei. Free neutron mass fractions will be only of order $\sim 10\%$, while free proton mass fractions will be even smaller in the mildly neutron-rich conditions expected in collpase. Therefore, with the rise of density as the collapse proceeds, the mean nuclear mass will become larger and larger. 

The neutrino-nucleus cross section in the coherent elastic scattering limit for the Standard Model neutral current is proportional to the {\it square} of the nuclear mass. Therefore, 
the trend of increasing nuclear mass eventually causes the core's opacity to neutrinos to become large enough for  neutrinos to be \lq\lq trapped.\rq\rq\   The large opacity causes trapping because neutrino mean free paths become less than the physical scale of the collapsing core and likewise mean diffusion (random walk) times become longer than the collapse timescale. 

Trapping sets in when the central density of the core is in excess of $\rho_{\rm trap} >{10}^{11}\,{\rm g}\,{\rm cm}^{-3}$. Somewhere between a density of ${10}^{13}\,{\rm g}\,{\rm cm}^{-3}$ and ${10}^{14}\,{\rm g}\,{\rm cm}^{-3}$, a series of phase transitions will take place where nuclei merge into sheets or tubes of material at nuclear density (the so-called \lq\lq pasta phase\rq\rq ). At higher density, these entities will merge into homogeneous nuclear matter.

The temperature during the infall epoch will likely not change drastically from a value $T\approx 2\,{\rm MeV}$. This is because heavy nuclei can store a large amount of energy in excited many-body states, thereby providing the medium with what amounts to a large specific heat. We will discuss the general composition and thermodynamic state for the stellar core and the prospects for entropy generation in section IV below.

Nuclear statistical equilibrium (NSE) obtains in the stellar collapse environment, so there may be a fair range of nuclear masses represented in the medium.  The spins and parities of these target nuclei could span a large range.  Nevertheless, the two spin cases we consider here in the elastic coherent scattering limit will serve as a guide to the behavior of the nuclear cross sections in general and, in particular, encompass the case of free nucleons.  Typical mean nuclear masses in the density regime where SUSY interactions could play a significant role are $A\approx 100$ to $200$, though the actual NSE distribution of nuclear abundances is sensitive to entropy, neutron excess, the history of electron capture reactions and neutrino transport \cite{Fuller82,hix}. 

A complicating issue in estimating neutrino-nucleus cross sections in the inelastic channel is that the target nuclei 
in stellar collapse will not be in their ground states. The large temperatures in the collapsing core, together with the large level densities that characterize heavy nuclei, imply that a typical nucleus will be in a state with an excitation energy of many tens of MeV above ground, possibly $\sim 100\,{\rm MeV}$. The mean nuclear excitation energy will be roughly the product of the number of nucleons excited above the nuclear Fermi surface (Fermi energy $\sim 40\,{\rm MeV}$), and the typical energy of each of these excited nucleons. The first of these quantities is $\approx aT$, where the level density parameter is $a\approx (A/8)\,{\rm MeV}^{-1}$ for nuclear mass $A$. If the nuclear level density is high enough, we can approximate the nucleons excited above the Fermi sea to be in plane wave states. In this limit, each excited nucleon has an energy of order the temperature, so that the mean excitation energy is $\langle E\rangle \approx a T^2$.

Neutrino-nucleon interactions in collapse will nearly always involve a nucleon inside a nucleus. The amplitudes for neutrino-nucleon interactions, whether mediated by  SM or non-standard processes, will therefore likely involve nontrivial medium effects on account of the dense environment where the nucleons and nuclei are found.   However, we can follow the usual procedure for Standard Model nuclear weak interactions in collapse and assume that nucleons in nuclei will have neutrino interaction properties broadly similar to those of bare nucleons, but restrict ourselves to processes that are kinematically allowed in the medium.  To this end, we derive neutrino-nuclear elastic scattering cross sections for nuclei and bare nucleons.  

We derive the cross sections in terms of a general low energy effective Lagrangian for neutrino-quark interactions.  We define what the general operators in this Lagrangian are for i) the Standard Model and ii) our SUSY model.  We then give the cross sections in terms of the general operators and comment on these specific cases.

Throughout, operators are implicitly evaluated at $x=0$. For
example, \( \bar q\g^\mu q\) stands for  \( \bar q(0)\g^\mu
  q(0)\). We take particle states to be relativistically normalized,
\begin{equation}
\vev{p'|p}=2E(2\pi)^3\delta^{(3)}(\vec p-\vec{p}\, '). \label{normalize}
\end{equation} 
We will use $k,k'$ for neutrino momenta and $p,p'$ for the momenta of nuclei.  Here, unprimed and primed variables are for incoming and outgoing particles, respectively.  We also recall here the invariant Mandelstam variables $s$, $t$ and $u$ (defined by $N(p)+\nu(k)\to N(p')+\nu(k')$)
\bea
s&\equiv&(p+k)^2=(p'+k')^2\\
t&\equiv&(p-p')^2=(k-k')^2 \\
u&\equiv&(p-k')^2=(k-p')^2,
\eea
with
\[
s+t+u=\sum_i m^2_i.
\]
In the rest frame of the nucleus $s$, $t$, and $u$ are
\bea 
s&=&M^2+2ME_\nu\\
t&=&\frac{-4ME_\nu^2\sin^2(\theta/2)}{M+2E_\nu\sin^2(\theta/2)} \label{t} \\
u&=&M^2-2E_\nu M+\frac{4ME_\nu^2\sin^2(\theta/2)}{M+2E_\nu\sin^2(\theta/2)},
\eea
where $\theta$ is the scattering angle, $M$ is the nuclear rest mass and the neutrino mass has been neglected.

\subsection{Effective Lagrangian}
We can write a general expression for the low energy effective Lagrangian for neutrino-quark interactions as
\beq
{\cal L}_{\rm eff}^{\rm Gen}=[\bar \nu^j\g^\mu\Pm \nu^i]
\sum_q[\bar q\g^\mu(\kappa^{(q)}_{Vij} +\kappa^{(q)}_{Aij}\g_5 )q],
\label{genlag}
\eeq
where the sum runs over all quarks.
For the Standard Model (SM) neutral current, the Lagrangian\cite{Peskin} has the usual current-current form
\begin{eqnarray}
\mathcal{L}_{\textrm{\scriptsize eff}}= 
\frac{G_F}{\sqrt{2}}\left [ \bar{\nu}_e\gamma^\mu(1-\gamma^5)\nu_e\right ] \times \nonumber \\ 
\times 2\left [\sum_{q_{\scriptscriptstyle{\rm L}},q_{\scriptscriptstyle{\rm R}}} 
\bar{q}\gamma_\mu\left (T^3 - \sin^2\theta_w Q\right )q \right ], \label{nclag}
\end{eqnarray}
where the sum runs over left and right handed helicity states of all quarks.
For this SM case, the operators in Eq. \parenref{genlag} are,
\bea
\kappa^{(q)}_{Vij}&=&\frac{G_F}{\sqrt2}(T^3-2\SW Q)\delta_{ij} \label{smv} \\ 
\kappa^{(q)}_{Aij}&=&-\frac{G_F}{\sqrt2}T^3\delta_{ij}.   \label{sma}
\eea
In Eq.s \parenref{smv} and \parenref{sma}, $T^3$ and $Q$ are the isospin and charge of quark $q$,
respectively. For example, we have $T^3=1/2(-1/2)$ and $Q=2/3(-1/3)$ for $q=u(d)$.

For the R-breaking SUSY Lagrangian (in particular, that given in Eq. \parenref{fclag1} but where the couplings are complex) the operators as defined in Eq. \parenref{genlag} are 
\bea
\kappa^{(d)}_{Vij}&=&\frac18\sum_k\left[
\frac{\lambda'_{i1k}(\lambda'_{j1k})^*}{m^2_{\tilde d^k_R}}
-\frac{\lambda'_{ik1}(\lambda'_{jk1})^*}{m^2_{\tilde d^k_L}}
\right] \label{kappaVsusy} \\
\kappa^{(d)}_{Aij}&=&\frac18\sum_k\left[
\frac{\lambda'_{i1k}(\lambda'_{j1k})^*}{m^2_{\tilde d^k_R}}
+\frac{\lambda'_{ik1}(\lambda'_{jk1})^*}{m^2_{\tilde d^k_L}}
\right]. \label{kappaAsusy}
\eea
In summary, the basic current-current structure of the low energy effective Lagrangians for the Standard Model flavor-conserving and SUSY flavor-changing neutrino interactions are identical. This affords a simple extension of the usual nuclear systematics of the Standard Model interactions to putative SUSY-inspired flavor changing reactions. It should be kept in mind that other non-SUSY flavor changing neutrino-quark interactions may have a completely different form with, consequently, nuclear response characteristics quite different from those presented in the following subsection.  However, our derivation of the coherent elastic cross sections in the subsequent subsections will apply to any flavor changing interactions that can be described by the Lagrangian in Eq. \parenref{genlag}. 

\subsection{General Nuclear Matrix Elements}
We first discuss general (i.e., not necessarily coherent elastic) neutrino scattering from nuclear states.  We will exploit here the similarity between the basic SUSY neutrino interaction operators and the Standard Model ones, as was discussed above. Again, we warn the reader that other, non-SUSY inspired flavor changing neutrino-quark interactions may have quite different nuclear operators.

We describe the $i$th excited state of a nucleus with $Z$ protons and $N$ neutrons with a general many-body state ket $\ket{\Psi_i(p)}$ which can be thought of as a superposition of Dirac spinors for individual nucleons.  Here the nuclear wave function $\Psi_i$ describes the full many-body nuclear state which, in the low energy nonrelativistic limit would have good quantum numbers corresponding to excitation energy $E_i$, angular momentum/parity $J_i^\pi$, and isospin $T_i$ with $z$-projection $T^3_i=(Z-N)/2$.  The kinematical bulk center of mass momentum of the state is labeled by $p$.

Of course, given the temperature and density conditions expected in stellar collapse as outlined above, and given the relatively modest neutrino energies expected to be encountered, we are not likely, at least initially at neutrino trapping, to require the full relativistic description of nucleons in the nuclei. Therefore, we will begin our discussion of the nuclear cross sections by considering general ({\it e.g.}, inelastic) neutrino scattering on nonrelativistic nuclei and nucleons. 

The small components of the full nuclear many-body wave functions/spinors can be taken to be linear superpositions of Slater determinants of nonrelativistic two-component single particle wave functions. These single particle states we envision to be computed in the usual independent particle model regime ({\it e.g.} shell model or RPA), as Hartree-Fock solutions with a specified two-body nucleon-nucleon potential. Likewise, the coefficients in the linear superposition of Slater determinants are presumed to come from an overall diagonalization procedure involving the residual interaction between nucleon quasi-particles in the nucleus.
 
The formal nuclear currents for general neutrino-nucleus scattering in the fully relativistic regime for either the Standard Model or SUSY-inspired neutrino interactions are:
\begin{eqnarray}
\label{relme1}
{\hat{V}}^\mu &=& {\bar\Psi}_f (p')\gamma^\mu{\Psi}_i(p) \\
\label{relme2}
{\hat{A}}^\mu &=& {\bar\Psi}_f (p')\gamma^\mu\gamma_5{\Psi}_i(p).
\end{eqnarray}
Here $\Psi_f$ is the many-body field operator describing a final nuclear state obtainable by the scattering channel being considered. This could differ from the initial nuclear state by a difference in any of the 
quantum numbers mentioned above.

In particular, a change in excitation energy between the initial and final nuclear states constitutes an inelastic neutrino interaction. This could, in turn, be classified as either of two types. It could be an endothermic transition, where $E_f > E_i$, and the scattering neutrino gives up energy to the nucleus. It could be an exothermic transition, where $E_f < E_i$ and the scattering neutrino picks up energy from the (excited) nucleus. The latter process is possible because, as discussed above, the target nuclei of interest in stellar collapse are expected to be in very highly excited states.  

If we take the nonrelativistic limit for the nucleon spinors in Eq.s\ (\ref{relme1}) and (\ref{relme2}), and contract with the leptonic current with an assumed negligible momentum transfer (the allowed approximation), then we obtain the usual Fermi and Gamow-Teller forms for the nuclear matrix elements, the absolute squares of which are
\begin{eqnarray}
\label{Fermi}
{\vert M^{\rm F}_{i f} \vert }^2 &=& {{1}\over{2 J_i+1}} 
\sum_{i, f}{{\vert {\langle\psi_f\vert \sum_n{\tau (n)}\vert\psi_i\rangle } \vert }^2} 
\\
\label{Gamow}
{\vert M^{\rm GT}_{i f} \vert}^2 &=& {{1}\over{2J_i+1}} 
\sum_{i, f}{{\vert {\langle\psi_f\vert \sum_n{\tau (n) {\vec{\sigma}}(n)}\vert\psi_i\rangle }  \vert }^2}. 
\end{eqnarray}
Here $\psi_i$ and $\psi_f$ are the initial and final two component nuclear many body wave functions which correspond to the nonrelativistic reductions of $\Psi_i$ and $\Psi_f$, respectively. In the above definitions we have averaged over initial and summed over final states for the particular transition $i\to f$. The sum on $n$ is a sum over nucleons in the nucleus and $\tau$ is an operator which depends on the process mediating the transition. 
The total matrix element squared between initial and final nuclear states $i$ and $f$ can be written
\begin{equation}
\label{mefinal}
{\vert M_{i f} \vert}^2={\vert D_{\rm F} \vert}^2{\vert M^{\rm F}_{i f} \vert}^2+{\vert D_{\rm GT} \vert}^2 {\vert M^{\rm GT}_{i f} \vert}^2, 
\end{equation}
where $D_{\rm F}$ and $D_{\rm GT}$ are coupling constants, each of which depends on the particular scattering mode, as does the operator $\tau$. 

For example, in the Standard Model charged current weak interaction we have $D_{\rm F}=C_V$, the vector coupling constant, while $D_{\rm GT} = C_A$, the axial vector coupling constant.  In this case $\tau$ is the isospin raising or lowering operator for individual nucleons, so that $\sum_n{\tau(n)} = T^\pm$, the overall isospin raising/lowering operator. The Fermi matrix element between members of an isospin muliplet in the Standard Model charged current case is simply 
\begin{eqnarray}
\label{fer11}
{\vert M^{\rm F}_{i f} \vert }^2 & = &  
{{\vert {\langle\psi_f\vert {T^\pm} \vert\psi_i\rangle }  \vert }^2} 
\\
\label{fer22}
& = & T\left( T+1\right) -T^3 \left( T^3\pm 1\right).
\end{eqnarray}
The selection rules implied by this are $\Delta J=0$, and no parity or isospin change.
By contrast, the Gamow-Teller operator for the Standard Model case, $\sum_n{ \tau(n) {\vec{\sigma}} }$, is a spatial vector and an isovector, so that the selection rules are $\Delta J=0,1$ (but no $0\to 0$), no parity change and $\Delta T=0,1$ (no $0\to 0$).
We discuss the Standard Model neutral current analogs of these interactions below in our presentation of the coherent elastic scattering case.
 
The SUSY flavor changing neutrino interactions are a different mode, with different values for the couplings and different meaning for the operator $\tau$. We shall denote the couplings in the SUSY flavor changing mode as $D_{\rm F}=C_V^{\rm SUSY}$ and $D_{\rm GT}=C_{\rm GT}^{\rm SUSY}$ (which can be expressed in terms of the unkown couplings, \lps). For the $d$-quark flavor changing neutrino scattering channels discussed above, the operator $\tau$ simply counts $d$-quarks. In this case then, $\tau(n)= 3/2+t_z$, where $t_z$ is the $z$-component of isospin for individual nucleons.  Note that this operator yields $\tau(n)=2$ for neutrons and $1$ for protons. The Fermi matrix element between members of an isospin multiplet is
\begin{equation}
\label{susyF}
{\vert M^{\rm F}_{i f} \vert}^2 = 2N+Z,
\end{equation}
with selection rules the same as those for the Standard Model Fermi operator. Likewise, the selection rules for the Gamow-Teller operator in this case are the same as in the Standard Model case.

In both the Standard Model and SUSY cases the form of the operators determines selection rules and the general distribution of strength with excitation energy. Note that the Fermi operator commutes with the nuclear Hamiltonian if we neglect the Coulomb potential as small compared to the strong interaction. On the other hand, the Gamow-Teller operator does not commute with the nuclear Hamiltonian. This is because of the spin dependence of the nuclear Hamiltonian. As a result, the \lq\lq Gamow-Teller strength\rq\rq\ associated with a given initial nuclear state in general will be spread throughout the other states of the nucleus that meet the selection rules. In practice, it is known that in the Standard Model Gamow-Teller case the strength may be significantly collected in a narrow resonance. We would expect this for the SUSY case as well, but the different dependence of the SUSY operator on isopsin probably implies that the resonance excitation energy centroids, widths, and overall strength will be different. 

For larger momentum transfer the operators and the corresponding matrix elements will depend on the momentum.  For SM weak interactions, the momentum transfer dependence of the one-body operators and, hence, all nuclear matrix elements is classified by a comparison of inverse momentum transfer to nuclear size $R$. The ``allowed'' regime corresponds to values of this parameter $\sqrt{t} R < 1$, while \lq\lq forbidden\rq\rq\ weak interactions have larger values.  For typical
pre-bounce collapse conditions, neutrino energies are $E_\nu\sim 50\, {\rm MeV}$, while nuclear masses are large $A \geq 100$ (with nuclear radii given by $R\approx 1.2A^{1/3}\, {\rm fm}$).  The high
Fermi levels of either the $e^-$ or $\nu_e$ sea (and sometimes the internal nuclear weak selection rules) dictate relatively low momentum transfer.  With modest momentum transfers, the allowed approximation may be reasonable for both the electron capture and $\nu_e$-capture channels of the SM charged current weak interactions.  (See Ref. \cite{Fuller82} for an in-depth discussion of these points.)

Flavor changing scatterings involving initial state $\nu_e$'s likely will be somewhat different than in the SM neutral current case for nuclei.  Momentum transfer in this channel could be larger on account of little or no blocking of final state $\nu_\mu$ and/or  $\nu_\tau$ phase space. This difference is accentuated for inelastic and endothermic neutrino scattering channels, where the nuclear final state is at a higher excitation energy than the initial state, and the neutrino final state has a consequently lower energy than the initial state. For flavor-preserving $\nu_e$ interactions, the final state $\nu_e$ is apt to be blocked by the degenerate $\nu_e$ sea unless its energy is larger than the $\nu_e$ Fermi level. This is less likely to be the case for flavor changing interactions because there are essentially no $\nu_\mu$ or $\nu_\tau$ species in the infalling supernova core, absent flavor changing neutral currents (FCNC's).

As the collapse proceeds and densities and neutrino energies rise, momentum dependence in the operators and matrix elements may become important.  This would tend to skew the effects we discuss in the Analysis section to the highest energy $\nu_e$'s if the SUSY couplings are small, but at the highest densities or larger coupling it will make little difference.  

\subsection{Spin-0 Coherent Elastic Nuclear Scattering}
We now consider elastic scattering where the initial and final nuclear states are the same.  We here designate the ket describing a spin-0 nuclear state with total momentum $p$ as $\ket{(Z,N),p}$. This state is some linear combination of the states for individual quarks and nucleons which make up the nucleus, as discussed above.     
We need the matrix element of the hadronic current in Eq. \parenref{genlag} with respect to such a spin-0 nuclear state.  The matrix elements for the vector and axial vector pieces expressed in terms of form factors are:
\bea
\me{(Z,N)p'}{\bar q\g^\mu q}{(Z,N)p}&=&f_q(t)(p+p')^\mu,  \label{ff1}\\
\me{(Z,N)p'}{\bar q\g^\mu\g_5 q}{(Z,N)p}&=&0,  \label{ff2}
\eea 
where $t$ is the momentum transfer squared defined in Eq. \parenref{t}.   
The matrix element of the axial vector current is zero because the 
initial and final states have the same parity and thus the left side of Eq. \parenref{ff2} is odd under parity, but the vectors  $p^\mu$ and $p'^\mu$ cannot be combined on the right hand side into an object that is odd under parity.

The form factor $f_q(0)$ is fixed. The conserved charge \footnote{Note that this charge is not the electric charge.} associated with the vector current is
\begin{equation}
Q_q=\int d^3 x \bar{q}\gamma^0 q  . \label{chrg}
\end{equation}
For example, this charge for the $u$ quark is 
\(
Q_u=\int d^3 x u^\dagger u .
\)
The spin-0 nuclear states given above are eigenstates of the charge operators defined by Eq. \parenref{chrg}, and in particular, these charges count number of quarks.  Therefore, setting $\mu=0$ 
in Eq. \parenref{ff1} and using Eq. \parenref{normalize}, we have $f_q(0)=N_q$, where $N_q$ is the number of $q$-quarks in the $(Z,N)$-nucleus:
\bea
f_u(0)&=&2Z+N   \\
f_d(0)&=&Z+2N.  
\eea

The precise momentum dependence of the form factor in Eq. \parenref{ff1} is not known of
course, but we may be able to get insights into this by considering
the SM case in the context of typical stellar collapse conditions, as
discussed above.  In accordance with the arguments made there about
the role of forbidden transitions in collapse, in what follows we will
take $f(t)\simeq f(0)$. As we will see below, this is well justified in
the case of nucleons, for which the form factors are precisely known
and the approximation of treating them as constant introduces an error
no larger than 3.0\%.
 
For the general low energy effective Lagrangian given in Eq. \parenref{genlag} 
in the channel $\nu_i+(Z,N)\to \nu_j+(Z,N)$ and/or  
$\bar\nu_i+(Z,N)\to \bar\nu_j+(Z,N)$ the differential cross section is 
\beq
\frac{d\sigma_{ij}}{dt}=\frac{1}{\pi}|\kappa^{d}_{Vij} f_d(t) + \kappa^{u}_{Vij} f_u(t)|^2\left[
1+\frac{ts}{(s-M^2)^2}\right]. \label{spin0diffcs}
\eeq
Integrating over squared momentum transfer $t$, the total cross section is
\bea
\label{spin01}
\sigma_{ij}&=&\frac{1}{2\pi}|\kappa^{d}_{Vij} f_d(0) + \kappa^{u}_{Vij} f_u(0)|^2
\frac{(s-M^2)^2}s\\
&\approx& \frac{2}{\pi}|\kappa^{d}_{Vij} f_d(0) + \kappa^{u}_{Vij} f_u(0)|^2
E_\nu^2 .  \label{spin0cs}
\eea
Note that for the case of the SM neutral current (with couplings given by Eq.s \ref{smv} and \ref{sma}) this result agrees with the cross sections given in Ref. \cite{T&Snuc}.
For the R-parity violating SUSY model, where we have only a $d$-quark interaction, the term $\kappa^u_{Vij}f_u$ is not present in the preceding equations.  
In particular, the cross sections for this case, with operators given by Eq. \parenref{kappaVsusy}, are  
\beq
\sigma_{ij} \approx \frac{2}{\pi}|\kappa^{d}_{Vij}|^2 f_d(0)^2 E_\nu^2 . \label{spin0cssusy}
\eeq
The results in Eq.s \parenref{spin0diffcs}-\parenref{spin0cs} would apply as well to any general neutrino flavor changing interactions that can be described by the Lagrangian in Eq. \parenref{genlag}.

\subsection{Spin-$\frac12$ Nuclei}
For a target nucleus with initial state angular momentum $J=1/2$, the matrix elements expressed in terms of form factors are:
\begin{widetext}
\bea
\me{(Z,N)p's'}{\bar q\g^\mu q}{(Z,N)ps}&=&
\bar u^{s'}(p')\left[H_1(t)\g^\mu+
\frac{H_2(t)}{2M}i(p-p')_\nu\sigma^{\mu\nu}\right] u^{s}(p), \label{1stline} \\
&=&
\bar u^{s'}(p')\left[(H_1(t)-H_2(t))\g^\mu+
\frac{H_2(t)}{2M}(p+p')^{\mu}\right] u^{s}(p), \label{2ndline} \\
\me{(Z,N)p's'}{\bar q\g^\mu\g_5 q}{(Z,N)ps}&=&
\bar u^{s'}(p')\left[G_1(t)\g^\mu
+\frac{G_2(t)}{2M}(p-p')^{\mu}\right] \gamma_5 u^{s}(p).
\eea
\end{widetext}
The form factors in this case are $H_1(t)$, $H_2(t)$, $G_1(t)$ and $G_2(t)$.
The states are labeled by their momentum $p$, spin $s$, and
$u^s(p)$ is a Dirac spinor satisfying the Dirac equation $(\g\cdot p-M)u^s(p)=0$. 
In going from Eq. \parenref{1stline} to Eq. \parenref{2ndline} we used the Gordon
identity. Eq. \parenref{1stline} is the standard form, but Eq. \parenref{2ndline} is
simpler for cross section computations. Hence, our results are given in
terms of the combination
\beq
\tilde H_1\equiv H_1-H_2.
\eeq
The form factors are implicitly  labeled by the flavor of the quark in the
current.  Note that from the Dirac equation and conservation of momentum we have
\begin{equation}
(p-p')_\mu\bar\nu\g^\mu\Pm\nu=2m_\nu\bar{\nu}\gamma_5\nu. \label{ignore}
\end{equation}
In computing the cross section, we will neglect neutrino masses $m_\nu$.  As a result of this approximation, $G_2$ will be irrelevent. 

Current conservation gives
\beq
\label{eq:h1norm}
H_1(0)=\cases{2Z+N & for $q=u$\cr Z+2N & for $q=d$}
\eeq
If we replace the vector current $\bar q\g^\mu q$ by the
electromagnetic current $J^\mu_{\rm em}=\frac23\bar u\g^\mu
u-\frac13\bar d\g^\mu d$, the form factors $H_{1,2}$  are replaced by the well known 
electromagnetic form factors $F_{1,2}$. In that case, we have more
information. In particular, we know in this case that the magnetic moment in units of $e/2M$ is $\mu=
F_1(0)-F_2(0)$. 

Moreover, for nucleons we have two more pieces of information. They
form an iso-doublet and the electric and magnetic factors,
\bea
G_E&\equiv& F_1-\frac{q^2}{4M^2}F_2\\
G_M&\equiv& F_1-F_2,
\eea
are known empirically to have  common $q^2$-dependence,
\begin{eqnarray}
G_{E,M}(q^2)=G_{E,M}(0)\tilde g(q^2), \\
\tilde g(q^2)=\frac1{(1-q^2/M_*^2)^2},
\end{eqnarray}
where $M_*^2\approx 0.71$~GeV$^2$. 

So, in the case of nucleons, we know precisely the form factors for
both the $\bar d\g^\mu d$ and the $\bar u\g^\mu u$ currents. We define $H_i^{(p,n)}$ by
\begin{widetext}
\bea
\me{n}{\bar d\g^\mu d}{n}=\me{p}{\bar u\g^\mu u}{p}&=&
\bar u^{s'}(p')\left[H_1^{(n)}(t)\g^\mu+
\frac{H_2^{(n)}(t)}{2M}i(p-p')_\nu\sigma^{\mu\nu}\right] u^{s}(p),\\
\me{p}{\bar d\g^\mu d}{p}=\me{n}{\bar u\g^\mu u}{n}&=&
\bar u^{s'}(p')\left[H_1^{(p)}(t)\g^\mu+
\frac{H_2^{(p)}(t)}{2M}i(p-p')_\nu\sigma^{\mu\nu}\right] u^{s}(p),
\eea
\end{widetext}
and, as before take $\tilde H_1=H_1-H_2$. We have used isospin
symmetry, which gives two relations among the four matrix
elements. Using the 
normalization conditions in Eq.~(\ref{eq:h1norm}), the
electromagnetic form factors and isospin, we solve 
for the individual quark currents. We find
\bea
\tilde H_1^{(n)}(t)&=&(2\mu_p+\mu_n)\tilde g(t)\approx(3.7)\tilde
g(t)\\
\tilde H_1^{(p)}(t)&=&(\mu_p+2\mu_n)\tilde g(t)\approx(-1.0)\tilde
g(t)\\
H_2^{(n)}(t)&=&(2-2\mu_p-\mu_n) g(t)\approx(-1.7)
g(t)\\
H_2^{(p)}(t)&=&(1-\mu_p-2\mu_n) g(t)\approx(2.0)
g(t) ,
\eea
where 
\beq
g(t)\equiv \frac{\tilde g(t)}{1-t/4M^2}.
\eeq
It is worth pointing out that $H_i^{(p,n)}(t)$ varies little over the
range of interest, that is, from $t_{\rm min}\approx -4E^2_{\nu,{\rm max}}\approx
-0.01$~GeV$^2$ to $t_{\rm max}=0$. Indeed, $(g(0)-g(t_{\rm
  min}))/g(0)=3.0\%$. To the extent that nucleon form factors are generic, the
approximation of treating nuclear form factors as
constant seems reasonable.

Now we turn our attention to the axial current. As mentioned after Eq. \parenref{ignore}, $G_2$ will not be needed.  Therefore, we only address $G_1$.
For protons we have \cite{filippone}
\beq
G_1(0)=\cases{0.78\pm0.03 & for $q=u$\cr -0.48\pm0.03 & for $q=d$
\cr -0.14\pm0.03 & for $q=s$}
\eeq
while for neutrons the same expressions apply but with $u\leftrightarrow d$ exchanged.
It should be noted that the $t\approx0$ form factor for the difference
\begin{widetext}
\beq
\me{p's'}{\bar u\g^\mu\g_5 u-\bar d\g^\mu\g_5 d}{ps}=
\bar u^{s'}(p')g_A\left[\g^\mu
+\frac{2M}t(p-p')^{\mu}\right] u^{s}(p).
\eeq
\end{widetext}
is very well known from neutron beta decay: $g_A=1.2670\pm0.0035$

The differential cross section for spin-$\frac12$ nuclei for a general Lagrangian which has only neutrino 
interactions with with $d$ quarks is,  
\begin{widetext}
\bea
\frac{d\sigma_{ij}}{dt}&=&\frac{1}{2\pi(s-M^2)^2}
\Bigg\{|\kappa^{d}_{Vij}|^2\left(\left[2H_2^2(1-t/4M^2)+4H_2\tilde
  H_1\right]\left[(s-M^2)^2+st\right]
+\tilde H_1^2\left[2(s-M^2)^2+2 s t+t^2\right]\right)\nonumber\\
& &\qquad +|\kappa^{d}_{Aij}|^2G_1^2\left[2(s-M^2)^2+2 s t+
               t^2-4M^2t\right]
\pm2{\rm Re}(\kappa^{d}_{Vij}\kappa^{d*}_{Aij})\tilde H_1G_1\left[2t(s-M^2)+t^2\right]
\Bigg\}.
\eea
The $\pm$ takes the upper sign for neutrino scattering and the lower sign for
antineutrino scattering. 
The corresponding total cross section, approximating $H(t)\approx H(0)$ is then
\bea
\sigma_{ij} &=& 
\frac{(s-M^2)^2}{6\pi s^3}\Bigg\{|\kappa^{d}_{Vij}|^2\left(\frac1{4M^2} H_2^2
s\left[(s-M^2)^2 + 12M^2
  s\right]+6s^2\tilde H_1H_2+\tilde H_1^2\left[4s^2-2sM^2+M^4\right]\right) \nonumber\\
& & \qquad + |\kappa^{d}_{Aij}|^2G_1^2\left[4s^2+4sM^2+M^4\right]
\mp2{\rm Re}(\kappa^{d}_{Vij}\kappa^{d*}_{Aij})\tilde H_1G_1(s-M^2)(M^2+2s)
\Bigg\}\\
&\approx& 
\frac{2E_\nu^2}{3\pi}\Bigg\{|\kappa^{d}_{Vij}|^2\left(3 H_2^2  +6
\tilde H_1H_2+ 3\tilde H_1^2\right)  + 9|\kappa^{d}_{Aij}|^2G_1^2\Bigg\}.
\eea
(The accuracy of these cross sections in the free nucleon channel may be improved by utilizing the well determined form factors of Eq.s 49 - 52.) 
Note that the cross term ($\kappa^{d}_{Vij}\kappa^{d*}_{Aij}$) is of order $E_\nu^3$
and hence was neglected in the last line. However, it should be kept in mind that this cross term may be important for generating asymmetries.

For $\kappa^d_{V,A}$ given by Eq.s \parenref{kappaVsusy} and \parenref{kappaAsusy}, the above equations give the differential and total cross section for the R-parity violating SUSY model.  The results for the SM are obtained by exchanging $\kappa^d_V \to (\kappa^d_V + \kappa^u_V)$ and $\kappa^d_A \to (\kappa^d_A + \kappa^u_A)$ everywhere in the above equations and using the appropriate values for the form factors (that is, taking account of whether they came from a $u$ quark or $d$ quark current and including cross terms between them).    
\end{widetext}

\section{Analysis: FCNC Effects in Core Collapse}
\subsection{The Supernova Model} \label{snmodel}
Gravitational collapse and the transformation of gravitational binding energy into (mostly) neutrinos and a small amount of outgoing kinetic energy and radiation is thought to be the process which powers supernovae of Type II, Ib, and Ic. The energy resident in the degenerate seas of neutrinos in these objects is huge, constituting some $10\%$ of the rest mass of the compact object produced in the collapse. In dramatic contrast, the energy of the explosion (optical plus outgoing kinetic energy) is only $1\%$ of this. We clearly see that even small changes in the energy or flavor content of the degenerate electron and $\nu_e$ reservoir may affect the physics of collapse and explosion.

The gravitational collapse phenomenon can be exquisitely sensitive to lepton number violating physics. This is because stellar collapse generates prodigious degenerate seas of electrons and electron neutrinos in weak equilibrium in a well-ordered, low entropy state. These electron lepton number fermion seas dominate the pressure and, hence, the dynamics during the collapse process. Flavor changing interactions can convert some of this degenerate electron lepton number into seas of neutrinos with mu and tau lepton numbers, thereby altering the Fermi levels of the electrons and $\nu_e$'s and, consequently, changing the pressure and, to a lesser extent, the entropy in the collapsing star. 

Better theoretical and observational insight into core collapse supernovae could allow them to become the ultimate laboratories for studying neutrino physics beyond the Standard Model. Conceivably, some day we could obtain constraints on or even discoveries of new physics, otherwise unobtainable in conventional accelerator experiments and terrestrial laboratories.
We are, unfortunately, not currently at this level of understanding. However, this is not to say that new neutrino physics could not significantly alter our picture for how stars collapse and explode.  

The current paradigm for the core collapse supernova explosion mechanism involves gravitational collapse halted by a \lq\lq bounce\rq\rq\ at nuclear or super-nuclear density. The bounce of an inner core is accompanied by the generation of an initially energetic shock wave at its boundary. This shock has its energy degraded by nuclear photo-disintegration of heavy nuclei as it traverses the outer core (the material laying between the inner core surface and the edge of the initial iron core). As a result, the shock is weakened and becomes a standing accretion shock, incapable of exploding the star. It is thought, however, that this shock subsequently is revived and strengthened by neutrino heating, perhaps aided by convective and hydrodynamic  processes\cite{mezza1,mezza2}. 

The progenitor star of such an event would have mass $M > 10\,{\rm M}_\odot$. This star would evolve over some millions of years through a succession of nuclear burning stages, eventually producing an \lq\lq onion skin\rq\rq\ structure of layers of fossil ashes of each core burning stage along with active burning shells at their boundaries. At the center would be a core composed of iron peak material in Nuclear Statistical Equilibrium (NSE). The central density of this core will be $\rho \sim {10}^{10}\,{\rm g}{\rm cm}^{-3}$, or $\rho_{10} \sim 1$, and the central temperature will be $T\sim1\,{\rm MeV}$. The electrons in these conditions are relativistically degenerate and they supply nearly all of the support pressure. The electron Fermi energy is $\mu_e \approx 11.1\,{\rm MeV}{( \rho_{10} Y_e)}^{1/3}$, where $Y_e$ is the electron fraction or net number of electrons per baryon.  Comparing $\mu_e$ to the temperature $T$ suggests that the entropy is low. A detailed accounting of all degrees of freedom yields an entropy per baryon $s \approx 1$ in units of Boltzmann's constant\cite{BBAL}.  

The silicon burning shell will add iron peak material to the core, until it exceeds the Chandrasekhar mass ($1.2\,{\rm M}_\odot$ to $1.6\,{\rm M}_\odot$ in this case), whereupon the core will go dynamically unstable and collapse at an appreciable fraction of the free fall rate. 
As the collapse proceeds, the baryons principally will reside in heavy nuclei, as dictated by the low entropy and as described in detail above in section III. Forcing the baryons to be confined to heavy nuclei implies that the baryonic pressure contribution during collapse is nearly negligible. 

During this collapse process, electron capture on protons which reside in these big nuclei
\begin{equation}
\label{ecap}
e^- + A(Z,N) \to A(Z-1,N+1) +\nu_e \label{ecap}
\end{equation}
and on the very few free protons present will lower the core electron fraction $Y_e$ and produce $\nu_e$'s. Electron capture on nuclei tends to increase the entropy because it leaves daughter nuclei in states that lie above the mean thermal excitation energy by $\sim 3\,{\rm MeV}$, a typical spin-orbit nuclear shell splitting for representative nuclei early in the collapse. The entropy increase via this process is modest and is nowhere near large enough to melt the nuclei.  Only completely melting the nuclei would allow the baryonic pressure to become significant.  In turn, this would require an entropy increase of at least three units of Boltzman's constant per 
baryon\cite{fmws,fmw}.  

The collapse is initially homologous. That is, the infall velocity of a fluid element is proportional to its radius from the center.  The collapse time scale is of order the free fall time, in particular, a few seconds. As the collapse proceeds and the pressure is reduced (because electron capture reduces $Y_e$), only a smaller, inner core can continue to collapse homologously. 
This inner core region of mass $0.6-0.8\msolar$ remains homologous, while the outer region of mass $0.7-0.9\msolar$ collapses supersonically.  

The edge of the homologous core is roughly the sonic point in the infall velocity field. In a sense, the inner core is causally self connected via pressure (sound) waves. Therefore, this portion of the core will bounce as a unit when the central density becomes large enough for the nuclei to merge and the pressure becomes dominated by the nonrelativistic baryonic component. This will occur, typically, at $3$ to $5$ times nuclear matter saturation density.  Another consequence of the causal structure of the inner core is that its mass is essentially that of an instantaneous Chandrasekhar mass,
 \begin{equation}
 \label{instant}
 M_{\rm hc} \approx 5.8 Y_e^2 \msolar . \label{hcmass}
 \end{equation}
 This is a good approximation because electron degeneracy pressure dominates.
 
 As outlined in section III above, the growth of large nuclei during core infall eventually causes the $\nu_e$'s produced by electron capture to become trapped in the core. This occurs when their random walk, or diffusion times become longer than the collapse time scale. The principal arbiter of trapping initially is coherent elastic neutral current neutrino scattering on the heavy nuclei. This process has a cross section proportional to the square of the nuclear mass and the square of the neutrino energy. As a consequence, at first only the higher energy $\nu_e$'s are trapped. Lower energy neutrinos can still escape and take lepton number and entropy out of the core. 
 
 Neutral current coherent neutrino scattering also is conservative.  It is only when non-conservative neutrino-electron scattering and neutrino-neutrino scattering opacity sources become significant, and when hot nucleus de-excitation into neutrino pairs operates that the $\nu_e$ distribution evolves toward a thermal, equilibrated Fermi-Dirac form.   In this case the core approaches weak, or beta equilibrium. At this point neutrinos are truly trapped and the evolution of the core from this point on is nearly adiabatic. Note that neutrino pair ($\nu_\alpha$, $\bar\nu_\alpha$ with $\alpha=$e,$\mu$,$\tau$) production via electron bremsstrahlung and plasmon decay processes during collapse are suppressed on account of the low entropy and the extreme electron degeneracy and high Fermi level $\mu_e$. As a result, during infall there may be relatively fewer neutrino types other than $\nu_e$ in the medium.
 
In weak equilibrium the electron capture and reverse $\nu_e$ capture reaction rates are equal and larger than the collapse rate. In essence, the build up of a degenerate $\nu_e$ sea eventually blocks further electron capture. Any process that opened holes in the $\nu_e$ distribution would inevitably lead to further electron capture and a lower $Y_e$ and, hence, a smaller homologous core mass at bounce, with the deleterious consequences outlined below. 
 
 In weak equilibrium the chemical potentials of leptons and nucleons are related through
 \begin{equation}
 \label{weakeq}
 \mu_e-\mu_{\nu_e} \approx \hat{\mu}+\delta m_{\rm np},
 \end{equation} 
 where $\hat{\mu}\equiv {\tilde\mu}_{\rm n}-{\tilde\mu}_{\rm p}$ is the difference of the kinetic chemical potentials (not including rest mass) of neutrons and protons and $\delta m_{\rm np}\approx 1.293\,{\rm MeV}$ is the neutron-proton rest mass difference. In NSE, the nucleon chemical potentials are the same inside and outside nuclei. Inside nuclei $\hat{\mu}$ can be interpreted as the difference of the neutron and proton nuclear Fermi levels. This rises as electron capture proceeds and the medium and the nuclei become more neutron-rich. The electron neutrino Fermi level (chemical potential) is roughly $\mu_{\nu_e}\approx 11.1\,{\rm MeV} {(2\rho_{10} Y_{\nu_e})}^{1/3}$, where the net number of $\nu_e$'s over $\bar\nu_e$'s per baryon is $Y_{\nu_e}$. 
 
Simulations of hydrostatic stellar evolution coupled with weak interaction rates suggest that the electron fraction at the onset of collapse is $Y_e^{\rm init} \approx 0.42$. Simple estimates of self consistent electron capture and nuclear equation of state issues yield a typical electron fraction at neutrino trapping of about $Y_e^{\rm trap} \approx 0.35$. Subsequently, lepton capture reactions redistribute this electron lepton number so that when weak equilibrium obtains shortly before core bounce $Y_e \approx 0.3$ and $Y_{\nu_e} \approx 0.05$, to give a trapped total electron lepton number fraction equal to $Y_e^{\rm trap}$.  
 
At a time of about $10-100$ milliseconds after the neutrinos are trapped and attain beta equilibrium, the inner region of the core reaches nuclear density and halts its collapse.   The supersonically-infalling outer core material bounces off the inner core and a shock wave forms at their boundary.  The shock's initial energy can be approximated as the kinetic energy of the infalling outer core material.  This kinetic energy, in turn, is approximately the gravitational potential energy of the inner core.   In terms of $Y_e$ in the core at bounce, the {\it initial} shock energy depends on the electron fraction roughly like \cite{Fuller82} 
\begin{equation}
\label{initsh}
E_{\rm shock}^{\rm init} \sim (Y_e)^{10/3}.
\end{equation}
The initial shock energy is $\sim\,{10}^{51}\,{\rm ergs}$ in the standard collapse model, a figure tantalizingly close to the energy of the shock observed to emerge through the progenitor star's envelope. However, entropy considerations again dictate nuclear physics realities that result in the vitiation of the initial shock strength.

All of the infall kinetic energy of the inner homologous core is converted to heat at core bounce. Furthermore, at core bounce all of the energy stored in the excited states of heavy nuclei (see section III) is returned to the general medium of homogeneous nuclear matter created from the merged nuclei. The upshot is a high temperature for the shocked core, $T\approx 20\,{\rm MeV}$ to $70\,{\rm MeV}$\cite{prakash}. 

The entropy jump across the shock front is about a factor of ten. In these hot, disordered conditions behind the shock, NSE favors disintegration of nuclei, so-called nuclear photo-disintegration. However, since each nucleon is bound in a nucleus by on average $8\,{\rm MeV}$, the shock loses energy as it propagates through the low entropy material in the outer core. In fact, the shock will lose $\sim {10}^{51}\,{\rm ergs}$ for each $0.1\,{\rm M}_\odot$ of outer core material traversed.
The shock propagates through the outer core and the material beyond.  Not surprisingly, simulations show that the shock stalls, or more accurately becomes a standing accretion shock, a few hundred km from the center of the core. This all happens within a few milliseconds after core bounce.   

In this condition the shock cannot effect an ejection of the outer stellar material to space.  That is, it cannot explode the star.  However, there is still a huge resevoir of neutrino energy in the core.  Some of this energy could be transported and deposited behind the shock, reviving it and leading to an eventual explosion.  

The higher entropy and high temperatures of the post-shock, post-bounce core (the hot proto-neutron star) now allow rapid production of neutrino pairs of all flavors. The neutrino and antineutrino seas for muon and tau neutrinos are approximately the same with no net lepton number in either one, but there remains an excess of $\nu_e$'s over $\bar{\nu}_e$'s on account of the trapped electron lepton number in the star.

There is rough equi-partition of energy among the six active neutrino species ($\nu_e$,$\bar\nu_e$,$\nu_\mu$,$\bar\nu_\mu$,$\nu_\tau$,$\bar\nu_\tau$) in the hot proto-neutron star.  These neutrino species diffuse out of the core on a random walk time scale $\tau_{\rm diff}^\nu\sim$ seconds, so that the energy luminosity in {\it each} of the six neutrino species is a few times ${10}^{52}\,{\rm ergs}\,{\rm s}^{-1}$ during the first half second or so and falls to $\sim {10}^{51}\,{\rm ergs}\,{\rm s}^{-1}$ some $10\,{\rm s}$ after bounce. These represent huge neutrino fluxes. In fact, the goal of a successful supernova explosion theory is to explain how $1\%$ of this energy can be pumped into the shock, either through direct heating of the material behind the shock, perhaps aided by convection, or through alleviation of some of the photo-disintegration burden on the shock.    
  
In summary, the initial iron core with a radius of order the size of the earth collapses to a proto-neutron star with a radius of $\sim 40\,{\rm km}$ in about a second, promptly liberating $\sim {10}^{52}\,{\rm ergs}$ of energy, $90\%$ of which resides in neutrino seas of all flavors, and $10\%$ of which, some ${10}^{51}\,{\rm ergs}$, comprises infall kinetic energy which becomes the initial shock energy. The shocked core, now a hot proto-neutron star, undergoes subsequent quasi-static contraction on a neutrino diffusion (random walk) time scale, $\tau_{\rm diff} \sim 3\,{\rm s}$. On this time scale the hot proto-neutron star will contract to a radius of about $10\,{\rm km}$ and release ${10}^{53}\,{\rm ergs}$ in gravitational binding energy. More than $99\%$ this energy will appear as neutrinos of all kinds. The basic points of this scenario have been confirmed by the neutrino-induced events detected from SN 1987A.     

Flavor changing neutral current (FCNC) neutrino processes can affect the above scenario for gravitational collapse and shock generation and propagation significantly. If electron neutrinos change flavor then holes are opened in their Fermi Dirac sea allowing electron capture to proceed.  Therefore, the number of electrons will be reduced and, hence, less pressure will be provided by electrons.  The mu and tau neutrinos created from electron capture plus FCNC processes will contribute some extra  pressure but, as a result of their much smaller Fermi levels (chemical potentials), this will not compensate for the reduction of electron pressure.  

A lower electron fraction and consequently lowered pressure on infall results in a double whammy. Not only is the initial shock energy reduced as Eq.\ (\ref{initsh}) shows, but the homologous core mass is reduced, leaving a larger outer core and a consequently greater nuclear photo-dissociation burden for the shock to overcome. In the popular current model for core collapse supernovae, this would greatly lessen the chance of getting a viable shock and an observationally acceptable explosion.

And there is another consequence of the operation of flavor changing neutrino processes during core infall. The FCNC processes could allow the post-bounce core to possess significant net mu and tau lepton numbers. This could, in turn, alter the hot proto-neutron star de-leptonization history through feedback on the nuclear equation of state and neutrino transport. This would also alter the flavor content of the emergent neutrino fluxes and therefore possibly change the supernova neutrino signal.

\subsection{Evaluation of Cross Sections}
In this section we use the cross sections derived previously to estimate the number of flavor changing (FC) scatterings which neutrinos might undergo at different densities during core collapse.  These estimates are not meant as accurate predictions for the actual number of scatterings because we do not account for feedback in the system.  Instead, these estimates are meant to illustrate the possible consequences of FCNC interactions and to suggest where a proper treatment through numerical simulation is warranted.  For large enough SUSY couplings, we could have enough flavor changing events during collapse and in the post bounce neutron star to result in a core evolution and shock generation/propagation scenario significantly different from that in the standard supernova model with only SM neutrino interactions. To illustrate these results, we limit our discussion to the case of spin zero nuclei with $Z=N$.  This case  is arguably overly simplistic, but it serves to illustrate how and at what level new flavor changing neutrino processes can be important.  

The cross sections we need are given in Eq.\ (\ref{spin0cs}) for the SM neutral current and in  Eq.\ (\ref{spin0cssusy}) for the SUSY model.  The SM cross section for the case of $Z=N$ can be written as
\begin{equation}
\sigma = \frac{\sigma_0}{4}(\sin^2\theta_W)^2 A^2 \left (\frac{E_\nu}{m_e}\right)^2 , \label{ncz=n}
\end{equation}
where $A=Z+N$ and 
\begin{equation}
\sigma_0\equiv\frac{4G_F^2m_e^2\hbar^2}{\pi c^2}\approx 1.76\times10^{-44}\textrm{cm}^2.
\end{equation}
Note that $G_F^2$ has been absorbed into $\sigma_0$.  The SUSY model cross section can likewise be written as
\begin{equation}
\sigma_{ij}=\frac{\sigma_0}{4}\frac{|G_{FC}^{ij}|^2}{G_F^2}[2N+Z]^2 \left (\frac{E_\nu}{m_e}\right)^2 
\label{fccs1}
\end{equation}
where we have defined
\begin{equation}
\frac{G_{FC}^{ij}}{\sqrt{2}}\equiv 
\frac18\sum_k\left[
\frac{\lambda'_{i1k}(\lambda'_{j1k})^*}{m^2_{\tilde d^k_R}}
-\frac{\lambda'_{ik1}(\lambda'_{jk1})^*}{m^2_{\tilde d^k_L}}
\right]  \label{Gfc}
\end{equation}
Note the similarity between between $G_{FC}^{ij}$ and the Fermi constant which is defined as
\begin{equation}
\frac{G_{F}}{\sqrt 2}\equiv\frac{g^2}{8 m_W^2}.
\end{equation}
Both are of the same form: a dimensionless coupling constant squared divided by a mass squared. It would be especially desirable to define a flavor changing SUSY coupling $G_{FC}=|\lambda'|/8m^2_{\tilde{q}}$. This could be factored out of Eq.s \parenref{kappaVsusy} and \parenref{kappaAsusy}, and likewise Eq. \parenref{Gfc}.  Then, any differences in the
\lps might be described by some underlying symmetry for the model.  However, as the coupling constants are unknown we will use  the notation of Eq. \parenref{Gfc} to stress that we are dealing with flavor changing neutrino interactions that behave like the SM interactions. 
  
To evaluate the SUSY cross sections we need values for the constants $G_{FC}^{ij}$ defined in Eq. \parenref{Gfc}.  
Values for $G_{FC}^{ij}$ are obtained by substituting values for individual $\lambda'$'s, or specific products of
$\lambda'$'s, with indices corresponding to the particular neutrino flavors being considered.  

The $\lambda'$ coupling constants are constrained from many different experiments\cite{Chemtob,Barbier,BGHsusy}.  As can be seen from Eq. \parenref{qrklag}, a particular $\lambda'$ with one set of indices serves as the coupling constant for several vertices.  Thus, several different processes can involve the same couplings and more than one experiment can be used to constrain a particular coupling.  Hence, in the literature, claimed constraints on the couplings can vary.   
Some constraints from Ref. \cite{Chemtob} are:
\begin{eqnarray}
|\lambda'_{11k}| < 2\times 10^{-2} \leftrightarrow 2.9\times 10^{-1} \tilde{d}_R^k \\
|\lambda'_{1k1}| < 4\times 10^{-2} \leftrightarrow 7.1\times 10^{-1}\tilde{q}_L^k \\
|\lambda'_{21k}| < 6\times 10^{-2} \leftrightarrow 1.5\times 10^{-1} \tilde{d}_R^k \\
|\lambda'_{2k1}| < 1.8\times 10^{-1} \tilde{d}_L^k \\ 
|\lambda'_{31k}| < 1.2\times 10^{-1} \tilde{d}_R^k
\end{eqnarray}
The notation $|\lambda'|< n\times \tilde{d}^k$ is shorthand for 
$|\lambda'|< n \times (m_{\tilde d ^k}/100 {\rm GeV})$.
We point out that these constraints are consistent with the constraints \footnote{Constraints in these papers are given in terms of a parameter $\epsilon$, not on products of $\lambda'$'s directly.  Constraints on products of $\lambda'$'s can be solved for in terms of $\epsilon$ assuming a squark mass of 100 GeV.   Note that these constraints are on specific products of $\lambda'$'s, not on individual $\lambda'$'s.  Note also there are notational inconsistencies in Ref. \cite{BPWsol} in Table I and the definition of the parameter $\epsilon$.} from references \cite{GMPsol,Rsol,BPWsol,GPsol,Bsol,KBsol,NRVfcncsn,MKfcncsn}.  
The $\lambda'$'s above can be combined to give $G_{FC}^{ij}$ for 
\begin{eqnarray*}
\nu_e\leftrightarrow\nu_\tau \\
\nu_e\leftrightarrow\nu_\mu \\
\nu_\tau\leftrightarrow\nu_\mu 
\end{eqnarray*}
flavor changing processes.  If we ignore phases and assume that no drastic cancellations occur in Eq. \parenref{Gfc}, then for the values of \lps given above, the upper limit on $G_{FC}^{ij}$ can range from 
$10^{-6}{\rm GeV}^{-2}$to $10^{-8}{\rm GeV}^{-2}$\footnote{Note that these values for the coefficients are also consistent with constraints for general FC neutrino interactions given in Ref. \cite{A&C}.}.  

As discussed above, the dominant neutrino scattering opacity source after trapping is coherent nuclear scattering.  The number of scatterings is then given approximately by the random walk expression
\begin{equation}
N_{\rm scatt}\approx\left (\frac{R}{\lambda_T}\right )^2, \label{nscatt}
\end{equation}
where $R$ is the radius of the core and $\lambda_T$ is the distance a neutrino travels before  
scattering off  target nucleus $T$, {\it i.e.}, the mean free path between such scattering events.  The neutrino mean free path after neutrino trapping and before the formation of the nuclear \lq\lq pasta phase\rq\rq\ is given by 
\begin{equation}
(\lambda_T)^{-1}\approx n_T\sigma_T \label{mfp},
\end{equation}
where $n_T$ is the number density of target nuclei $T$, and $\sigma_T$ is the cross section for coherent scattering off such nuclei.  
The number of flavor changing (FC) scatterings $N_{\rm scatt}^{\rm FC}$that neutrinos undergo after trapping can be estimated from the number of neutral current (NC) scatterings.  Substituting Eq. \parenref{mfp} into Eq.\ \parenref{nscatt} and taking the ratio $N_{\rm scatt}^{\rm FC}/N_{\rm scatt}^{\rm NC}$ gives
\begin{equation}
N_{\rm scatt}^{\rm FC}\approx\left (\frac{\sigma_A^{\rm FC}}{\sigma_A^{\rm NC}}\right )^2N_{\rm scatt}^{\rm NC}, \label{nfcscatt1}
\end{equation}
where $\sigma_A^{\rm FC}$ and $\sigma_A^{\rm NC}$ are the FC and NC scattering cross sections on a nuclues of mass number $A$.
As alluded to above, to illustrate the effect of FC scatterings we only need to consider neutrino scattering off nuclei with $Z=N$ and so we use the cross sections given by Eq.s \parenref{ncz=n} and \parenref{fccs1}.  Since $(2N+Z)^2\sim A^2$, Eq. \parenref{nfcscatt1} with these cross sections becomes
\begin{equation}
\label{numscat}
N_{\rm scatt}^{\rm FC} \approx \left ( \frac{G_{FC}^2}{G_F^2\sin^2\theta_w} \right )^2 N_{\rm scatt}^{\rm NC}.
\end{equation} 
Taking the appropriate value of the Weinberg angle from experiment gives $\sin^2\theta_w=0.23$ which, together with the Fermi constant $G_{\rm F}=1.166\times 10^{-5}{\rm GeV}^{-2}$, allows Eq.\ (\ref{numscat}) to become
\begin{equation}
N_{\rm scatt}^{\rm FC} \approx 1\times 10^{21}[G_{FC}\  {\rm GeV}^2]^4 N_{\rm scatt}^{\rm NC} .
\label{nfcscatt3}
\end{equation}
With this expression we can calculate the number of FC scatterings as a function of the number of NC scattering for various values of $G_{FC}^{ij}$.  
We note that similar estimates for the number of scatterings for general flavor changing neutrino interactions could be obtained in this same way once they have been cast into the nuclear physics framework discussed in section III and their cross sections are known.

To get the number of NC scatterings we use Eq.s \parenref{nscatt} and \parenref{mfp}.  We take the cross section in Eq. \parenref{ncz=n} with a mean neutrino energy $E_\nu$ figured from the density in the core.  Likewise $n_A$, the number density of target nuclei of mass $A$, can be computed from the core density.   The mean free path is  \cite{S&T}
\begin{equation}
(\lambda_A^{\rm NC})^{-1}\simeq4\times 10^{-5} \rho_{12}^{5/3}{\rm cm}^{-1},
\end{equation}
where $\rho_{12}=\rho/10^{12}{\rm g}\,{\rm cm}^{-3}$.  

We estimate the number of FC scattering from Eq. \parenref{nfcscatt3} and calculate the mean free path for neutrinos to undergo an FC scattering using Eq. \parenref{nscatt}.  For these calcuations we approximate the radius of the core as $50\,{\rm km}$ over the whole range of densities we consider. Keep in mind, however, that the core radius will be some $45\,{\rm km}$ immediately after core bounce, and will shrink over several neutrino diffusion times $\tau_{\rm diff}^\nu$ to $\approx 10\,{\rm km}$.  
The rate of FC scatterings is obtained here by dividing the speed of light by the mean free path. 

We do these calculations for a range of densities covering the post neutrino-trapping infall epoch of collapse and the post bounce hot proto-neutron star.  We have used the cross sections in Eq.s \parenref{ncz=n} and \parenref{fccs1} over the entire range of these densities. This is not strictly valid at larger densitites where, as discussed in section III, nuclei do not exist and where neutrino opacities may be dominated by scattering on collective modes in nuclear matter or different phases of quark matter which can exist at high density.  However, coherent neutrino scattering effects with quarks can still occur during these phases \cite{reddy}.  Hence, our procedure may give a crude estimate for the number of scatterings during these phases,  and the level at which FCNC couplings may be large enough to be of significance.  

For our calculations, we will use $G_{FC}$ values of $10^{-6}{\rm GeV}^{-2}$, $10^{-7}{\rm GeV}^{-2}$,  
$10^{-8}{\rm GeV}^{-2}$  and $10^{-9}{\rm GeV}^{-2}$. The case where we take
$G_{FC} = 10^{-9}{\rm GeV}^{-2}$ covers the situation where at least one of the \lps is an order of magnitude smaller than the smallest value suggested by current experimental bounds.
We give our results in Tables \ref{-7}-\ref{-9}. These results are meant only as a guide to the order of magnitude of the number of FC neutrino scatterings and, hence, to the level of FC effects in the core.   

The estimated total number of scatterings for a given density is not meant to be the total number for all the densities at that order of magnitude. Rather, they are a guide to the total number of scatterings during each of several time intervals (which typically are a few milliseconds) as the central density of the core passes through each regime.  Additionally, results given for 
higher densities were not computed with values corresponding to a core in which FC processes had been operative.  In other words, we have neglected feedback here and any semblance of a self consistent collapse history when FC effects are appreciable.  In our conclusions we discuss more about possible additional feedback the FC processes could have on the model and stress again that the numbers in these tables are merely a guide.

\subsection{Analysis}
The numbers provided in the tables above give insight into the range of possible physical effects of the FC interactions on collapse and shock generation/propagation physics and the rough level of FC coupling where these effects can be expected to be important.  They suggest that for some plausible as yet unconstrained ranges of these couplings, $\nu_e$'s can change flavor. Comparing the rates for 
electron capture (essentially given by the inverse of the weak equilibration time scale) and possible FC processes,  we see that it is likely that the former process will be fast compared to the latter one. In this limiting case, every time a hole is opened in the $\nu_e$ sea by an FC interaction, it is immediately filled by a $\nu_e$ created by electron capture. This approximation is used and discussed in Ref. \cite{fmws}. 

Another limiting case is one in which the FC processes are fast and equilibrium among flavors is achieved and all the Fermi levels of the neutrino seas are the same, though not necessarily at the maximum level.  This case likely would have to correspond to larger values of $G_{\rm FC}$.  

FC interactions will tend to change $\nu_e$ to $\nu_{\mu,\tau}$ and thereby build up the muon and tau neutrino seas, but not {\it vice versa}.  This is because the $\nu_e$ sea is already in weak equilibrium and possesses a rather high Fermi level. Because of this significant $\nu_e$ Fermi level, $\nu_{\mu,\tau} \to \nu_e$ tends to be blocked.  Of course, it is possible that a hole in the $\nu_e$ sea could get filled by a $\nu_e$ created in a $\nu_{\mu,\tau}$ FC interaction before it gets filled by a $\nu_e$ created via electron capture.  However,
as $\sigma_A^{e^- {\rm cap}}$ is larger than $\sigma^{\rm FC}_A$, holes are more likely to be filled by electron capture.  The main point is that after trapping, in the limiting case where electron capture rates can be regarded as fast comparted to FC rates, the level of the $\nu_e$ sea remains the same as it would be absent FC interactions.  

We can quantify the effect of the FC interactions for this limiting case by using the approximation that the level of the $\nu_e$ sea remains the same.  The lepton fraction (total lepton number of all kinds per baryon) in the core is given by 
\begin{equation}
Y_L=Y_e+Y_{\nu_e}+Y_{\nu_\mu}+Y_{\nu_\tau},
\end{equation}
where the net number fraction relative to baryons is defined as discussed in subsection \ref{snmodel} by, 
\begin{equation}
Y_f \equiv \frac{n_f-n_{\bar{f}}}{n_b} \label{Yf}.
\end{equation}
We can evaluate this at neutrino trapping and again at some time past trapping.
At trapping, there is no excess of $\nu_{\mu,\tau}$'s over $\bar{\nu}_{\mu,\tau}$'s because these species have all been produced as neutrino-antineutrino pairs via thermal emission processes.  Therefore, by Eq. \parenref{Yf} we have 
\begin{equation}
Y_{\nu_\mu}^{\rm trap} = Y_{\nu_\tau}^{\rm trap} = 0.
\end{equation}
At later times, if our FC interactions exist and are taking place, then\ $Y_{\nu_\mu,\nu_\tau}$ increases while $Y_{\nu_e}$ remains the same and $Y_e$ decreases.  Therefore, the post-trapping $\nu_e$ fraction is fixed in this limit,
\begin{equation}
Y_{\nu_e}^{\rm pt} = Y_{\nu_e}^{\rm trap}.
\end{equation}
As discussed above above, after trapping the total number of relativistic leptons is fixed if we neglect neutrino radiation from the surface of the collapsing core.  Therefore, $Y_L$ is constant in this limit and we have the relation,
\begin{equation}
Y_e^{\rm trap}+Y_{\nu_e}^{\rm trap} =
Y_e^{\rm pt}+Y_{\nu_e}^{\rm trap}+Y_{\nu_\mu}^{\rm pt}+Y_{\nu_\tau}^{\rm pt} .
\end{equation}
In other words, the change in electron fraction is given by
\begin{equation}
\Delta Y_e = -\left ( Y_{\nu_\mu}^{\rm pt}+Y_{\nu_\tau}^{\rm pt} \right ).
\end{equation}
The effect of FC interactions essentially is to convert $\Delta Y_e$ to $Y_{\nu_\mu,\nu_\tau}$.

\begin{figure}
\resizebox{8cm}{!}{\includegraphics{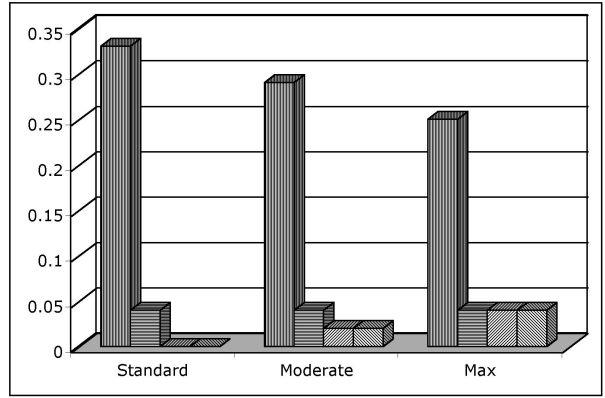}}
\caption{Possible scenarios for the effect of FC interactions on the levels of the various fermion seas. The bin with verticle bars represntes $Y_e$, that with horizontal bars represents $Y_{\nu_e}$, and those with diagonal bars represent $Y_{\nu_{\mu,\tau}}$. ``Standard'' refers to the standard collapse model in weak equilibrium. ``Max'' refers to the maximal effects of FCNCs as discussed in the text, while ``moderate'' refers to an intermediate case.}
\label{figlevels}
\end{figure}

We do not give specific values for $\Delta Y_e$ because we are not equipped to account for feedback.  We can however point out limiting cases as illustrated in Figure \ref{figlevels}.  The extreme limiting case on infall is 
\begin{equation}
Y_{\nu_\mu}=Y_{\nu_\tau}=Y_{\nu_e}^{\rm trapp}.  \label{maxconv}
\end{equation}
The muon and tau neutrino seas are built up to the maximum level before bounce.  This limit is actually consistent with the current constraints on the $\lambda'$'s.  For the result of Eq.\ \parenref{maxconv} to obtain, we require a total number of electron neutrinos equal to $2\times N_{\nu_e}^{\rm trapp}$ to change flavor.  Given that the rate of FC interactions for the value of $G_{\rm FC} = 1\times 10^{-7}{\rm GeV}^{-2}$ is of order 10 scattering per millisecond for $\rho_{12}=10$
we have more than enough room for this situation.  For tighter constraints on the $\lambda'$'s which give $G_{\rm FC}$ of order $10^{-8}{\rm GeV}^{-2}$, $\mathcal{R}\sim 10$ scatt/ms at $\rho_{12}=100$ with enough time prior to bounce to build up the $\nu_{\mu,\tau}$ seas making the extreme limiting case possible. 

For this case, we would expect $\Delta Y_e \approx -2Y_{\nu_e} \approx -0.1$. This is a large effect, on the same order as the change in electron fraction from the introduction of new electron capture 
physics \cite{rphstlk}. It would reduce the initial shock energy by a factor of $\sim 3$ and increase the nuclear photo-disintegration burden on the shock by a substantial amount. This would, in turn, imply that the shock would stall at a smaller radius, likely decreasing the efficacy of neutrino re-heating.  However, although these effects argue in the standard picture against a viable explosion, caution is called for and detailed shock propogation physics must be employed. (Again see Ref. \cite{rphstlk} on this issue where it was pointed out that consistent electron capture physics on infall and during shock reheating is necessary to gauge the viablity of the shock.)
Ultimately, we do not understand where explosions come from, so that constraints on SUSY or other non-SM FCNC interactions cannot be drawn with any degree of confidence yet.  Clearly, to accurately assess the effects of FCNC's on collapse/explosion, our cross sections should be incorporated into detailed simulations with full hydrodynamics, nuclear equation of state, and Boltzman neutrino transport.

\section{Conclusions}

We have shown that new flavor changing neutral current (FCNC) neutrino interactions could have significant effects in the physics of core collapse supernovae. In particular, we have outlined how these interactions can both reduce the initial shock energy and increase the amount of material that the shock must traverse and photo-disintegrate. Both of these effects go in the direction of causing the shock to weaken and, therefore, to stall and become a standing accretion shock at a smaller radius. This likely would greatly decrease the prospects for obtaining a viable shock and explosion in the current paradigm for the core collapse supernova explosion mechanism. 

Another consequence of a significant level of FCNC neutrino interactions during collapse would be the creation of net mu and tau lepton numbers in the post-bounce hot proto-neutron star core. This differs from the standard supernova collapse scenario, where these lepton numbers would be zero. This could affect neutrino transport and the equation of state in the de-leptonizing core. It could also alter the neutrino radiation emergent from the core and, thereby change the neutrino signal expected from a core collapse event.  

However, as emphasized at the end of the last section, only detailed collapse simulations with coupled nuclear physics, hydrodynamics, and neutrino transport with our FCNC cross sections can adequately ascertain what happens in the supernova.  We have demonstrated that there are as yet unconstrained ranges of R-parity violating SUSY FCNC parameters that could yield alterations in weak interaction physics and therefore the FCNC's are worthy of incorportation into these simulations.  Likewise, other kinds of flavor changing (non-SUSY) neutrino-nucleus interactions would produce alterations in lepton physics during collapse identical in a qualitative sense to what we have considered here.  

There are also effects which we have not analyzed.  For example, anti-neutrino flavor changing due to these interactions would have similar cross sections, and in particular $A^2$-like dependence.  Another effect is the potential for entropy increase due to flavor changing inelastic scattering.  That is, entropy increase due to high energy $\nu_e$'s scattering into low energy $\nu_{\mu,\tau}$'s (because these low energy states are not blocked) leaving nuclei in excited states and thus heating the system.  These effects treated properly in a simulation might lead to further interesting results.  

Additional effects we have not analyzed include FC interactions involving the charged leptons. Note that the last two terms in the Lagrangian in Eq. (\ref{qrklag}) and their Hermitian conjugates can mediate $e^-\rightarrow \mu^-$. During the early stages of the post-neutrino trapping infall regime this process will not be operative because the electron Fermi energy $\mu_e$ will be less than the muon rest mass ($m_\mu \approx 106\,{\rm MeV}$). However, near core bounce and post core bounce,  where the density is of order nuclear saturation density, it is possible to have $\mu_e > m_\mu$.  In this case we would expect $e^-\rightarrow \mu^-$, and for large enough FC coupling equilibrium between these species would result.  This would augment the trends discussed above which stem from neutrino FC reactions.  Namely, with large scale conversion in the channel $e^-\rightarrow \mu^-$ we would expect a further lowering of $Y_e$ and increase in net muon lepton number.  

We have outlined the nuclear physics of FCNC reactions.  We have found that,  similarly to the SM NC case, there is a coherent elastic scattering neutrino FC channel with a cross section that scales as $(2N+Z)^2$.  Given the low entropy, and consequently the large neutron-rich nuclei  expected in the infall epoch of stellar collapse, this factor can be sizable.   

As we have seen, the square of the cross section comes into the calculation for number of scatterings.  Since nuclei would be even more neutron rich because of increased $e^-$ capture if FCNCs were taken into account, the factor of 2 in $2N+Z$ could be important.  We neglected this factor of 2 (factor of 4 in the cross sections) and related feedback in our estimates for numbers of scatterings.  If we included this factor, the numbers of scatterings could increase by up to a factor of 16.  

Our considerations for the results of neutrino flavor changing apply to any general FC interactions which give large enough numbers of FC scatterings.  Large cross sections would likely arise from neutrino quark interactions that allowed for coherent scattering amplifications.  For a general interaction with neutrinos on electrons, for example, the number of scatterings would likely not be large .  Note that our R-parity violating SUSY model has neutrino-electron flavor changing interactions.  In particular, the second and third terms in the first line of Eq. \parenref{leplag} (which originates from the superpotential term in Eq. \ref{lepsf}) lead to neutrino-electron flavor changing.  The products of $\lambda$'s that appear here again result in ranges of values for a low energy effective coefficient similar to the ranges we used for $G_{FC}$.

In the case where FCNCs are the result of our R-parity violating SUSY model, even couplings an order of magnitude and possibly more below experimental bounds could have significant effects on the infall epoch of stellar collapse. This naturally begs the question of whether we can use the considerations presented here to provide constraints on these couplings or even, conceivably, to suggest signatures of non-SM neutrino interactions in, for example, the supernova neutrino signal. 

We argue that the current lack of understanding of the origin of explosions in core collapse supernovae precludes constraints and signatures which can be treated with the same confidence as those obtained from current accelerator-based and terrestrial laboratory experiments. However, there is intense interest in core collapse supernovae and new insights are being gained from large scale simulations with state of the art numerical hydrodynamics, nuclear physics, and neutrino transport \cite{janka}. Ultimately, a better understanding of these events may result and this, in turn, could allow the considerations discussed in this paper to be turned into legitimate constraints and/or signatures. The prospects for this would be greatly increased by the detection of a neutrino burst signal from a collapse event in our galaxy.

On the other hand, the experimental discovery of supersymmetry would necessitate a new assesment of the stellar collapse problem.  The discovery of R-parity violating SUSY in particular would be a strong motivation for FCNCs.  In this case the stellar collapse effects of FCNCs outlined in this paper would become potentially vital (depending on the strength of coupling constants) in understanding the core collapse supernova phenomenon. 

We would like to point out a cautionary historical tale for the supernova modeling community. In the 1970's the discovery of SM neutral currents in the laboratory (and their theoretical prediction) completely altered the model for core collapse at the time. Likewise, the existence of new FCNC interactions could significantly alter the current model for collapse and explosion of massive stars. 

\section{Acknowledgments}
The work of P.S.A. and G.M.F. was supported in part by NSF grant PHY-00-99499 and
the TSI collaboration's DoE SciDAC grant at UCSD;
while that of B. G. was supported in part by DoE grant DE-FG03-97ER40546 at UCSD. The authors thank the Institute for Nuclear Theory at the University of Washington for hospitality and assistance. We would like to thank A. Friedland, R. Hix, C. Lunardini, B. Messer, A. Mezzacappa, S. Reddy, and B. Wecht for useful discussions, and Georg Raffelt for pointing out the high density FCNC conversion of electrons to muons.

\clearpage

\clearpage

\begin{table*}
\begin{tabular}{|c|c|c|c|} \hline
$\lambda'_{ijk}\tilde{d}^j_L\bar{d}^k_R\nu^i_L$ & $\lambda'_{ijk}(\tilde{d}^j_L)^*\bar{\nu}^i_Ld^k_R$
&$\lambda'_{ijk}(\tilde{d}^k_R)^*\overline{\nu^c}^i_Ld^j_L$  & $\lambda'_{ijk}\tilde{d}^k_R\bar{d}^j_L\nu^{ci}_L$ \\
\hline 
\resizebox{4.2cm}{!}{\includegraphics{vertex1.epsf}} & \resizebox{4.2cm}{!}{\includegraphics{vertex2.epsf}} 
& \resizebox{4.2cm}{!}{\includegraphics{vertex3.epsf}}  & \resizebox{4.2cm}{!}{\includegraphics{vertex4.epsf}} \\
\hline
\end{tabular}
\caption{Vertices of interactions.  The first and third terms appear in Eq. (\ref{qrklag}) while the second and fourth are  Hermitian conjugates of the first and third. The indices are assigned as $e,d=1$; $\mu,s=2$; $\tau,b=3$.  The couplings here are taken to be real.}
\label{table 1}
\end{table*} 

\clearpage

\begin{table}
\begin{tabular}{|c|c|c|c|c|}\hline 
$\rho_{12}$ & $N_{\rm scatt}^{\rm NC}$ & $N_{\rm scatt}^{\rm FC}$ &
$\lambda_A^{\rm FC} = R/(N^{\rm FC}_{\rm scatt})^{1/2}$  &  $\mathcal{R}= c/\lambda $ \\
$10^{12}{\rm g}/{\rm cm}^3$ & &  & m & scatt/ms \\
\hline
$1$       &  $4\times 10^4$      & $4\times 10^1$       & $8\times10^3 $ & $40 $ \\
\hline
$10$     &  $9\times 10^7$      & $9\times 10^4$       & $2\times10^2 $ & $2\times 10^3$ \\
\hline
$100$   &  $2\times 10^{11}$ & $2\times 10^8$      & $4$                       & $8\times 10^4$ \\
\hline
$1000$ & $4\times 10^{14}$ & $4\times 10^{11}$  & $8\times 10^{-2}$ & $4\times 10^6$ \\
\hline
\end{tabular}
\caption{Values for FC interactions for $G_{FC}=1\times 10^{-6}{\rm GeV}^{-2}$ Mean free paths are given in meters, rates are given as scatterings per millisecond.}
\label{-6}
\end{table}

\begin{table}
\begin{tabular}{|c|c|c|c|c|}\hline 
$\rho_{12}$ & $N_{\rm scatt}^{\rm NC}$ & $N_{\rm scatt}^{\rm FC}$ &
$\lambda_A^{\rm FC} = R/(N^{\rm FC}_{\rm scatt})^{1/2}$  &  $\mathcal{R}= c/\lambda $ \\
$10^{12}{\rm g}/{\rm cm}^3$ & &  & m & scatt/ms \\
\hline
$1$       &  $4\times 10^4$      & $4\times 10^{-3}$ & $8\times10^5 $ & $.4 $ \\
\hline
$10$     &  $9\times 10^7$      & $9$             & $2\times10^4 $ & $20$ \\
\hline
$100$   &  $2\times 10^{11}$ & $2\times 10^4$    & $4\times 10^2$ & $8\times 10^2$ \\
\hline
$1000$ & $4\times 10^{14}$ & $4\times 10^7$     & $8$                     & $4\times 10^4$ \\
\hline
\end{tabular}
\caption{Values for FC interactions for $G_{FC}=1\times 10^{-7}{\rm GeV}^{-2}$ Mean free paths are given in meters, rates are given as scatterings per millisecond.}
\label{-7}
\end{table}

\begin{table}
\begin{tabular}{|c|c|c|c|c|}\hline 
$\rho_{12}$ & $N_{\rm scatt}^{\rm NC}$ & $N_{\rm scatt}^{\rm FC}$ &
$\lambda_A^{\rm FC} = R/(N^{\rm FC}_{\rm scatt})^{1/2}$  &  $\mathcal{R}= c/\lambda $ \\
$10^{12}{\rm g}/{\rm cm}^3$ & &  & m & scatt/ms \\
\hline
$1$       &  $4\times 10^4$      & $4\times 10^{-7}$  & $8\times10^7$ & $4\times 10^{-3} $ \\
\hline
$10$     &  $9\times 10^7$      & $9\times 10^{-4}$ & $2\times10^6$  & $2\times 10^{-1}$ \\
\hline
$100$   &  $2\times 10^{11}$ & $2$                          & $4\times 10^4$ & $8$ \\
\hline
$1000$ & $4\times 10^{14}$ & $4\times 10^3$      & $8\times 10^2$ & $4\times 10^2$ \\
\hline
\end{tabular}
\caption{Values for FC interactions for $G_{FC}=1\times 10^{-8}{\rm GeV}^{-2}$ Mean free paths are given in meters, rates are given as scatterings per millisecond.}
\label{-8}
\end{table}

\begin{table}
\begin{tabular}{|c|c|c|c|c|}\hline 
$\rho_{12}$ & $N_{\rm scatt}^{\rm NC}$ & $N_{\rm scatt}^{\rm FC}$ &
$\lambda_A^{\rm FC} = R/(N^{\rm FC}_{\rm scatt})^{1/2}$  &  $\mathcal{R}= c/\lambda $ \\
$10^{12}{\rm g}/{\rm cm}^3$ & &  & m & scatt/ms \\
\hline
$1$       &  $4\times 10^4$      & $4\times 10^{-11}$ & $8\times10^9$ & $4\times 10^{-5}$ \\
\hline
$10$     &  $9\times 10^7$      & $9\times 10^{-8}$  & $2\times10^8$ & $2\times 10^{-3}$ \\
\hline
$100$   &  $2\times 10^{11}$ & $2\times 10^{-2}$  & $4\times 10^6$ & $8\times 10^{-2}$ \\
\hline
$1000$ & $4\times 10^{14}$ & $4\times 10^{-1}$  & $8\times 10^4$ & $4$ \\
\hline
\end{tabular}
\caption{Values for FC interactions for $G_{FC}=1\times 10^{-9}{\rm GeV}^{-2}$ Mean free paths are given in meters, rates are given as scatterings per millisecond.}
\label{-9}
\end{table}


\clearpage

\begin{thebibliography}{99}

\bibitem{GMPsol}M.~M.~Guzzo, A.~Masiero and S.T.~Petcov, Phys. Lett. B \textbf{260}, 154 (1991).

\bibitem{Rsol}E.~Roulet, Phys. Rev. \textbf{D 44}, R935 (1991).

\bibitem{BPWsol}V.~Barger, R.~J.~N.~Phillips and K.~Whisnant, Phys. Rev. D. \textbf{44}, 1629 (1991).

\bibitem{GPsol}M.~M.~Guzzo and S.~T.~Petcov, Phys. Lett. B \textbf{271}, 172 (1991).

\bibitem{Bsol}S.~Bergmann, Nucl. Phys. \textbf{B515}, 363 (1998).

\bibitem{KBsol}P.~I.~Krastev and J.~N.~Bahcall, hep-ph/9703267.

%\bibitem{BKfcsn}S. Bergmann and A. Kagan, Nuc. Phys. \textbf{B538}, 368 (1999).

%\bibitem{Vfosn}J. W. F. Valle, Phys. Lett. B \textbf{199} 432 (1987).

\bibitem{NRVfcncsn}H.~Nunokawa, A.~Rossi and J.~W.~F.~Valle, Phys. Rev. D \textbf{56}, 1704 (1997).

\bibitem{MKfcncsn}S.~W.~Mansour and T.~K.~Kuo, Phys. Rev. D \textbf{58} 013012 (1998).

\bibitem{FLMMfcncsn}G.~L.~Fogli, E.~Lisi, A.~Mirizzi and D.~Montanino, Phys. Rev. D \textbf{66}, 013009 (2002).

\bibitem{MSW}L.~Wolfenstein, Phys. Rev. D \textbf{17}, 2369 (1978); \textbf{20}, 2634 (1979); S.~P.~Mikheyev
and A.~Y.~Smirnov, Nuovo Cimento \textbf{9C}, 17 (1986).

\bibitem{Chemtob} M.~Chemtob, hep-ph/0406029

\bibitem{Barbier}R.~Barbier \emph{et al.}, hep-ph/9810232.

\bibitem{fmws}G.~M.~Fuller, R.~W.~Mayle, J.~R.~Wilson and D.~N.~Schramm, Astrophys. J. {\bf 322}, 795 (1987).

\bibitem{fmw}G.~M.~Fuller, R.~W.~Mayle, J.~R.~Wilson, Astrophys. J. {\bf 332}, 826 (1988).

\bibitem{susy}H.~P.~Nilles, Phys. Rep. \textbf{110}, 1(1984); H.~E.~Haber and G.~L.~Kane, Phys. Rep. \textbf{117}, 75 (1985).

\bibitem{WBsusy}J.~Wess and J.~Bagger Supersymmetry and Supergravity, 2nd Ed. (Princeton University Press, Princeton, 1992).

\bibitem{BLsusy}D.~Balin and A.~Love, Supersymmetric Gauge Field Theory and String Theory,  (Institute of Physics Publishing, London 
1994).

\bibitem{Msusy}S.~P.~Martin, hep-ph/9709356, v3: 1999 (extended version of a contribution to the book \emph{Perspectives in Supersymmetry}, edited by G.~L.~Kane (World Scientific, Singapore, 1998).

\bibitem{JGmssm}For a self contained review of the details of constructing  the MSSM see Apendix A:  G.~Jungman, M.~Kamionkowski and K.~Griest, Phys. Rep. \textbf{267}, 195 (1996).

\bibitem{rparity}L.~J.~Hall and M.~Suzuki, Nuc. Phys. {\bf B231}, 419 (1984).

\bibitem{BGHsusy}V.~Barger, G.~F.~Giudice and T.~Han, Phys. Rev. D \textbf{40}, 2987 (1989).

\bibitem{Herczeg}P.~Herczeg, Phys. Rev. D {\bf 61} 095010, 2000.
%e-Print Archive: hep-ph/9912495

%\bibitem{Fnuc}D. Z. Freedman, Phys. Rev. D \textbf{9}, 1389 (1974).

%\bibitem{FSTnuc}D. Z. Freedman, D. N. Schramm and D. L. Tubbs, Ann. Rev. Nucl. Sci. \textbf{27}, 167 (1977).

\bibitem{BBAL} H. ~A.~ Bethe, G.~E.~Brown, J.~Applegate, and J.~Lattimer, Nucl. Phys. {\bf A324}, 487 (1979).

\bibitem{Fuller82}G.~M.~Fuller, Astrophys. J. {\bf 252}, 741 (1982). 

\bibitem{hix}W.~R.~Hix {\it et al.}, Phys. Rev. Lett. {\bf 91}, 201102 (2003).

\bibitem{Peskin}M.~E.~Peskin and D.~V.~Schroeder, An Introduction to Quantum Field Theory, (Perseus Books Publishing, L.L.C., Cambridge, 1995).

\bibitem{T&Snuc}D.~L.~Tubbs and D.~N.~Schramm, Astrophys. J. \textbf{201}, 467 (1975).

\bibitem{filippone}B.~W.~Filippone and X.~Ji, Adv. Nucl. Phys. {\bf 26}, 1 (2001).
%hep-ph/0101224
% eq (56))

\bibitem{mezza1}J.~Blondin, A.~Mezzacappa and C.~DeMarino, Astrophys. J. {\bf 584}, 971 (2003).

\bibitem{mezza2}A.~Mezzacappa {\it et al.}, Phys. Rev. Lett. {\bf 86}, 1935 (2001).

\bibitem{prakash}M.~Prakash, J.~M.~Lattimer, J.~A.~Pons, A.~W.~Steiner and S.~Reddy, Lect. Notes Phys. {\bf 578}, 364 (2001).

\bibitem{A&C}A.~Friedland, C.~Lunardini and C.~Pena-Garay, Phys. Lett. {\bf B594}, 347 (2004).

\bibitem{S&T}S.~L.~Shapiro and S.~A.~Teukolsky, Black Holes, White Dwarfs, and Neutron Stars, (John Wiley and Sons, Inc., New York, 1983).

\bibitem{reddy}S.~Reddy, G.~Bertsch and M.~Prakash, Phys. Lett. {\bf B475}, 1 (2000).

\bibitem{rphstlk}W.~R.~Hix, in Proceedings of the workshop Open Issues in Understanding Core Collapse Supernovae, Seattle, 2004, edited by A. Mezzacappa. 

\bibitem{janka}H.~T.~Janka {\it et al.}, 
%R.~Buras, K.~Kifonidis, M.~Rampp, and T.~Plewa, 
%Explosion Mechanisms of Massive Stars
in {\it Core Collapse of Massive Stars}, edited by C.~L.~Fryer,  
(Kluwer, Dordrecht, 2003).
%Astrophysics and Space Sciences Library {\bf 302} p. 66 (2003) 

\end{thebibliography}
\end{document}